\documentclass[10pt]{article}
\usepackage{bbm}
\usepackage[final]{graphics}
\usepackage{amsmath}
\usepackage{amsfonts,amsbsy}
\usepackage{amssymb}
\usepackage{color}
\definecolor{linkcolor}{rgb}{0.4,0.1,0.1}
\definecolor{bibcolor}{rgb}{0.4,0.1,0.1}

\def\empile#1\over#2{\mathrel{\mathop{\kern 0pt#1}\limits_{#2}}}
\def\bs{\boldsymbol}

\newcommand{\slv}{\raise.15ex\hbox{$/$}\kern-.53em\hbox{$v$}}
\newcommand{\slF}{\raise.15ex\hbox{$/$}\kern-.53em\hbox{$F$}}
\newcommand{\slL}{\raise.15ex\hbox{$/$}\kern-.53em\hbox{$L$}}
\newcommand{\slP}{\raise.15ex\hbox{$/$}\kern-.53em\hbox{$P$}}
\newcommand{\slp}{\raise.15ex\hbox{$/$}\kern-.53em\hbox{$p$}}
\newcommand{\slq}{\raise.15ex\hbox{$/$}\kern-.53em\hbox{$q$}}
\newcommand{\slR}{\raise.15ex\hbox{$/$}\kern-.53em\hbox{$R$}}
\newcommand{\slQ}{\raise.15ex\hbox{$/$}\kern-.53em\hbox{$Q$}}
\newcommand{\slK}{\raise.15ex\hbox{$/$}\kern-.53em\hbox{$K$}}
\newcommand{\slk}{\raise.15ex\hbox{$/$}\kern-.53em\hbox{$k$}}
\newcommand{\slD}{\raise.15ex\hbox{$/$}\kern-.73em\hbox{$D$}}
\newcommand{\slC}{\raise.15ex\hbox{$/$}\kern-.53em\hbox{$C$}}
\newcommand{\slA}{\raise.15ex\hbox{$/$}\kern-.53em\hbox{$A$}}
\newcommand{\slSigma}{\raise.15ex\hbox{$/$}\kern-.53em\hbox{$\Sigma$}}
\newcommand{\slpartial}{\raise.15ex\hbox{$/$}\kern-.53em\hbox{$\partial$}}
\newcommand{\slcalP}{\raise.15ex\hbox{$/$}\kern-.63em\hbox{$\cal P$}}

\def\P{{\boldsymbol P}}

\def\Q{{\boldsymbol Q}}
\def\l{{\boldsymbol l}}
\def\k{{\boldsymbol k}}

\def\x{{\boldsymbol x}}
\def\y{{\boldsymbol y}}
\def\X{{\boldsymbol X}}

\def\v{{\boldsymbol v}}

\def\u{{\boldsymbol u}}

\begin{document}

\title{\bf %
Role of quantum fluctuations\\ 
in a system with strong fields:\\
Spectral properties and Thermalization}
\author{Thomas Epelbaum, Fran\c cois Gelis}
\maketitle
\begin{center}
Institut de Physique Th\'eorique (URA 2306 du CNRS)\\
  CEA/DSM/Saclay, 91191 Gif-sur-Yvette Cedex, France
\end{center}

\begin{abstract}
  In a previous work [arXiv:1009.4363], we have studied the evolution
  of a scalar field with a quartic coupling, driven by a classical
  source that initializes it to a non-perturbatively large value. At
  leading order in the coupling, the evolution of this system is given
  by classical solutions of the field equation of motion. However,
  this system is subject to a parametric resonance that leads to
  secular divergences in higher order corrections to physical
  observables. We have proposed a scheme that resums all the leading
  secular terms: this resummation leads to finite results at all
  times, and we have observed also that it makes the pressure tensor
  of the system relax to its equilibrium value.

  In the present paper, we continue the study of this system by
  looking at finer details of its dynamics. We first compute its
  spectral function at various stages of the evolution, and we observe
  that after a fairly short transient time there are well defined
  massive quasi-particles. We then consider the time evolution of the
  momentum distribution of these quasi-particles, and we show that
  after a stage dominated by the parametric resonance, this
  distribution slowly evolves to an equilibrium
  distribution. Interestingly, this distribution develops a transient
  chemical potential, signalling the fact that number changing
  processes are much slower than the elastic ones.
\end{abstract}

\section{Introduction}
\label{sec:intro}
The issue of thermalization is a very challenging question in the
theory of heavy ion collisions. On the one hand, a lot of
circumstantial evidence suggests that the matter formed in these
collisions behaves like a nearly perfect fluid at very short times
after the collision (this is based on a comparison of flow
measurements with predictions of hydrodynamical models -- see
\cite{RomatR1} for instance). This fact is commonly interpreted as a
sign of the fast thermalization of the quark-gluon matter produced in
nucleus-nucleus collisions\footnote{Note however than some systems
  with disordered field configurations may also exhibit an anomalously
  low viscosity \cite{AsakaBM1,AsakaBM2}.}. On the other hand,
justifying this thermalization from first principles in QCD has proven
to be very difficult, and to this date the question has not been
satisfactorily answered.

The main issue is the following: a semi-hard momentum scale --the
saturation momentum $Q_s$-- is dynamically generated by the non-linear
gluon interactions that occur in nuclear collisions at high energy
\cite{GriboLR1,MuellQ1,BlaizM2}. This scale controls the so-called
gluon saturation phenomenon, that alters the gluonic content of a
nucleus at high energy, but it is also this scale that controls the
final state interactions that occur after the collision, at least
initially. However, the saturation scale $Q_s$ increases with the
energy of the collision, and therefore the relevant QCD coupling
effectively decreases. It thus seems that the relevant degrees of
freedom are rather weakly coupled in collisions at very high energies,
rendering difficult the obtention of short thermalization times
\cite{BaierMSS2}.

To compute the initial gluon production in heavy ion collisions at
high energy, we have at our disposal an effective theory --the Color
Glass Condensate
\cite{McLerV1,McLerV2,McLerV3,IancuLM3,IancuV1,GelisIJV1,Lappi6}--
that consistently takes into account the non-linear saturation
effects. In this effective theory, the degrees of freedom that
populate the light-cone wave-function of a nucleus are divided in two
classes, according to their longitudinal momentum. Partons with a
longitudinal momentum above a certain cutoff $\Lambda$ are
approximated by a collection of static (because they appear highly
boosted in the frame of the slower partons) color sources of density
$\rho_a(\x_\perp)$, while the partons with longitudinal momentum below
$\Lambda$ are treated as standard gauge fields\footnote{At the leading
  order of this description, the gluons are dominant over the
  quarks. Only the former are taken into account in the description.}
$A^\mu(x)$. In this effective theory, the sources $\rho_a$ are
stochastic variables (reflecting the fact that we do not know the
precise arrangement of the fast partons at the time of the collision),
with a probability distribution $W[\rho]$.

From the above description of the CGC, it may seem that observables
could depend on the unphysical cutoff $\Lambda$, that has been
introduced by hand as a separation scale between the fast and the slow
partons. This worry becomes manifest when one computes loop corrections
in this framework: they usually contain logarithms of the cutoff
$\Lambda$. Fortunately, one can show that for sufficiently inclusive
observables in the collision of two nuclei, these logarithms are
universal in the following sense: (i) they do not depend on the
observable under consideration, but are instead properties of the
initial state, and (ii) each logarithm of $\Lambda$ can be assigned to
one of the two projectiles (meaning that its coefficient does not
depend on the nature of the second projectile). This universality
allows to absorb all these logarithms into a redefinition of the
distributions $W[\rho]$ of the two projectiles. After this is done,
these distributions depend on the cutoff $\Lambda$ (in such a way as
to cancel the $\Lambda$ dependence that arises from loop corrections),
and their changes under variations of the cutoff are controlled by the
JIMWLK evolution equation \cite{JalilKMW1,JalilKLW1,JalilKLW2,JalilKLW3,JalilKLW4,IancuLM1,IancuLM2,FerreILM1,GelisLV3,GelisLV4,GelisLV5}.

The Color Glass Condensate, supplemented with the JIMWLK equation,
therefore provides a consistent framework for computing observables
relevant to heavy ion collisions, such as the spectrum
\cite{KrasnV2,KrasnV3,KrasnNV1,KrasnNV2,KrasnNV4,Lappi1,Lappi7} of
produced gluons or components of the energy-momentum tensor
\cite{KrasnV1}. However, these computations have also shown that the
gluonic matter produced in these collisions is rather far from local
equilibrium \cite{LappiM1}. In fact, its pressure tensor is highly
anisotropic, with two positive transverse pressures and a negative
longitudinal pressure. Such a form of the pressure tensor is arguably
too far from the equilibrium form to justify the applicability of
hydrodynamics (even with viscous corrections).

This computation, a leading order calculation in $\alpha_s$ improved
by the resummation of all the leading logarithmic contributions
discussed above, is however not the end of the story. It has been
noted in various situations
\cite{RomatV1,RomatV3,FujiiI1,FujiiII1,FukusG1,BiroGMT1,HeinzHLMM1,KunihMOST1}
that the classical solutions of Yang-Mills equations that constitute
the LO answer in the CGC approach are unstable under small
perturbations. Practically, this means that loop corrections contain
secular divergences --i.e. terms that have an unbounded growth with
time-- and that the perturbative expansion breaks down after a certain
time (this time can be estimated to be of the order of $Q_s^{-1}$, up
to logarithms). An additional resummation is therefore necessary in
order to circumvent this problem and to obtain expressions that remain
well behaved at all times.

Interestingly, the structure of the NLO (1-loop) corrections to
inclusive observables in the CGC formalism suggests a natural
resummation scheme, that in a loose sense amounts to exponentiating
the 1-loop result. Such a resummation first appeared in the context of
inflation in cosmology \cite{Son1,KhlebT1}. In the context of the
color glass condensate and heavy ion collisions, it was sketched in
\cite{GelisLV2} in the case of gauge fields, and more thoroughly
justified\footnote{This resummation was also derived in a completely
  different approach in \cite{FukusGM1}. The full spectrum of initial
  fluctuations in Yang-Mills theory in the Fock-Schwinger gauge has
  been recently derived in \cite{DusliGV1}.}  in the case of scalar
fields in \cite{DusliEGV1}. The reason why a $\phi^4$ scalar field
theory in four dimensions is an interesting playground for testing
these ideas is that this theory, like Yang-Mills theory at the
classical level, is scale invariant, and that its classical field
configurations are also subject to an instability (due to parametric
resonance\footnote{The problem of secular divergences and parametric
  resonance has been addressed in a number of other approaches in
  Quantum Field Theory, including perturbation theory (in the regime
  where the resonant modes are still small) and the 2-Particle
  Irreducible
  resummation~\cite{Serre3,BergeS3,BergeBS1,AartsB1,AartsB2,AartsLT1}.}). The
resummation we advocated in \cite{DusliEGV1} can be formulated in two
equivalent ways: its derivation leads first to a rather formal
expression, in which the resummation is expressed as the action of the
exponential of a certain operator on the LO result. This form is not
practical for a numerical implementation, but is equivalent to
averaging the LO order result (which is expressed in terms of a
classical field) over a Gaussian ensemble of initial conditions for
the classical field. We then showed that the proposed resummation
indeed leads to observables that remain finite at all times, therefore
solving the original problem of secular divergences. Moreover, by
studying the time evolution of the pressure tensor, we showed that
after the resummation, the pressure converges towards its equilibrium
value in a fairly short time --something that would not happen with
the unresummed expression.

In this paper, we pursue the study of this system of scalar fields, in
order to elucidate the microscopic state of the system. In particular,
we would like to know whether the system is thermalized when its
pressure has relaxed to its equilibrium value, or whether on the
contrary one could have an equilibrium-like pressure tensor while the
system is still far from equilibrium\footnote{In the 2PI approach to
  the relaxation of a linear sigma model, it has already been observed
  that an equation of state can be obtained much earlier than the
  complete thermalization of the system \cite{BergeBW1}.}. A natural
quantity to study in order to address this question is the occupation
number in the system, and its time evolution. However, even before
computing the occupation number, it is interesting to ask whether the
system can be described in terms of quasi-particles (this is not
trivial: although the system is weakly coupled, it is also very dense,
and strong collective effects may render the quasi-particles
completely unstable). We start with a reminder of the model and of the
main results of \cite{DusliEGV1} in the section \ref{sec:setup}; then
we compute in the section \ref{sec:toy:spectral} the spectral density
of the system, after having justified that it can be resummed in the
same way as the energy-momentum tensor. In the section \ref{sec:occ},
we continue our study with the occupation number. We compute it as a
function of momentum up to large times, and identify several stages in
its time evolution. In particular, we discuss the possible occurrence
of turbulence and Bose-Einstein condensation. From the occupation
number, we perform in the section \ref{sec:qp-mass} several tests of
the quasi-particle picture (e.g. compare the measured mass of the
quasi-particles, with its value at 1-loop, including medium effects)
and compute how the entropy of the system evolves with time. A summary
and further discussion can be found in the section
\ref{sec:t-evol}. In the appendix \ref{sec:classical}, we discuss some
aspects of the classical dynamical system to which the quantum field
theory is equivalent in our resummation scheme. Some basic properties
of the Liouville equation are recalled in the appendix
\ref{app:liouville}.

\section{Setup of the problem - Summary of previous results}
\label{sec:setup}
\subsection{Model and resummation of secular terms}
In \cite{DusliEGV1}, we considered a real scalar field theory,
with quartic self-interactions, and coupled to an external source. The
Lagrangian of this model reads
\begin{equation}
{\cal L}\equiv \frac{1}{2}(\partial_\mu\phi)(\partial^\mu\phi)-\underbrace{\frac{g^2}{4!}\phi^4}_{V(\phi)}+J\phi\; .
\label{eq:L}
\end{equation}
In this model, the source $J$ is nonzero only at negative times, and
its purpose is to initialize the fields to a configuration that has a
large expectation value at $t=0$. At positive times, $J=0$ and the
fields evolve by themselves.

In  \cite{DusliEGV1},  we  have  studied  the time  evolution  of  the
energy-momentum tensor of the system. At leading order (tree level) in
the  coupling $g^2$,  it is  simply  the energy-momentum  tensor of  a
classical field:
\begin{eqnarray}
T^{\mu\nu}_{_{\rm LO}}(x)
&=&
\partial^\mu\varphi\partial^\nu\varphi
-
g^{\mu\nu}\,\Big[\frac{1}{2}(\partial_\alpha\varphi)^2-\frac{g^2}{4!}\varphi^4\Big]\; ,
\nonumber\\
\square \varphi +\frac{g^2}{3!}\varphi^3&=&J\; ,\quad 
\lim_{x^0\to -\infty}\varphi(x^0,\x)=0\; .
\label{eq:LO}
\end{eqnarray}
However, at NLO (1 loop) we observed that $T^{\mu\nu}$ is plagued by
secular divergences that originate from an instability of the above
classical solution, due to parametric resonance. In order to cure
this problem, we developed a resummation scheme that collects the
leading secular terms at each order of the expansion in $g^2$. Our
resummed expression reads
\begin{equation}
T^{\mu\nu}_{\rm resum}(x)
\!\equiv\!
\exp\!\Big[
\int \!\!d^3\u\, \beta\cdot{\mathbbm T}_\u
+\frac{1}{2}\!\int\! d^3\u d^3\v\!\int\!\frac{d^3\k}{(2\pi)^3 2k}
[a_{+\k}\cdot{\mathbbm T}_\u][a_{-\k}\cdot{\mathbbm T}_\v]
\Big]
T^{\mu\nu}_{_{\rm LO}}(x)
\, .
\label{eq:sum}
\end{equation}
In this formula, ${\mathbbm T}_\u$ is the generator of shifts of the
classical field on the $x^0=0$ hypersurface, defined formally
by
\begin{equation}
  a\cdot{\mathbbm T}_\u
  \equiv
  a(0,\u)\frac{\delta}{\delta\varphi_0(\u)}
  +
  \dot{a}(0,\u)\frac{\delta}{\delta\dot\varphi_0(\u)}\; ,
\end{equation}
where $\varphi_0(\u)$ and $\dot\varphi_0(\u)$ are the values of the
field and its first time derivative on the surface $x^0=0$.  This
shift operator possesses a very useful property: if $a(x)$ is a small
perturbation on top of a classical field $\varphi(x)$, one has
\begin{equation}
a(x)=\int d^3\u\;[a\cdot{\mathbbm T}_\u]\;\varphi(x)\; .
\label{eq:shift}
\end{equation}
The fields $a_{\pm\k}$ are small perturbations propagating on top of
the classical field $\varphi$, that are plane waves at
$x^0\to-\infty$, and $\beta$ is the 1-loop correction to $\varphi$,
\begin{eqnarray}
&&\Big[\square+V^{\prime\prime}(\varphi)\Big]a_{\pm\k}=0\; ,\qquad
\lim_{x^0\to-\infty}a_{\pm\k}(x)=e^{\pm ik\cdot x}\; ,\nonumber\\
&&\Big[\square+V^{\prime\prime}(\varphi)\Big]\beta = 
-\frac{1}{2}V^{\prime\prime\prime}(\varphi)
\int\frac{d^3\k}{(2\pi)^3 2k}\;a_{-\k}a_{+\k}\; ,\qquad
\lim_{x^0\to-\infty}\beta(x)=0\; .
\label{eq:fluctuations}
\end{eqnarray}
(See \cite{DusliEGV1} for more details.)  A crucial result is that
eq.~(\ref{eq:sum}) is equivalent to a functional integration over
Gaussian fluctuations of the classical field at $x^0=0$,
\begin{eqnarray}
T_{\rm resum}^{\mu\nu}
=
\int [D\alpha(\x) D\dot\alpha(\x)]\,F[\alpha,\dot\alpha]\;
T^{\mu\nu}_{_{\rm LO}}[\varphi_0+\beta+\alpha]\; ,
\label{eq:sum1}
\end{eqnarray}
where $T^{\mu\nu}_{_{\rm LO}}[\varphi_0+\beta+\alpha]$ denotes the LO
energy-momentum tensor evaluated with a {\sl classical field} whose
initial condition at $x^0=0$ is $\varphi_0+\beta+\alpha$ (and likewise
for the first time derivative). The distribution
$F[\alpha,\dot{\alpha}]$ is Gaussian in $\alpha(\x)$ and
$\dot\alpha(\x)$, with 2-point correlations given by
\begin{eqnarray}
\big<\alpha(\x)\alpha(\y)\big>&=&
\int\frac{d^3\k}{(2\pi)^3 2k}\;a_{+\k}(0,\x)a_{-\k}(0,\y)\; ,
\nonumber\\
\big<\dot\alpha(\x)\dot\alpha(\y)\big>&=&
\int\frac{d^3\k}{(2\pi)^3 2k}\;\dot{a}_{+\k}(0,\x)\dot{a}_{-\k}(0,\y)\; .
\label{eq:gaussian}
\end{eqnarray}
We refer the reader to \cite{DusliEGV1} for more details about the
model and its practical lattice implementation.

\subsection{Relaxation of the pressure}
\begin{figure}[htbp]
  \begin{center}
    \resizebox*{8cm}{!}{\rotatebox{-90}{\includegraphics{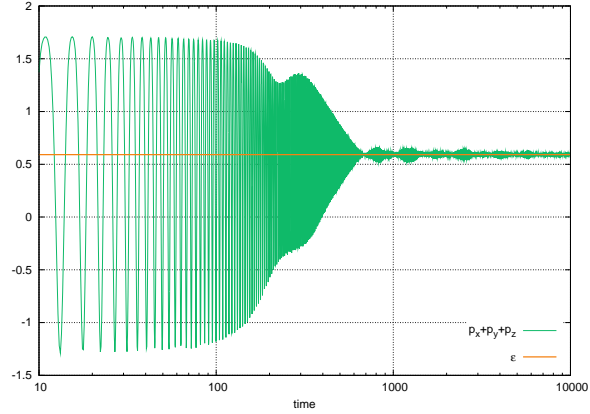}}}
  \end{center}
  \caption{\label{fig:eos}Relaxation of the pressure tensor towards
    the equilibrium value. All the numerical results presented in this
    paper have been obtained with a coupling constant $g=1$.}
\end{figure}
The main result of \cite{DusliEGV1} is that the fluctuations resummed
in the formula (\ref{eq:sum1}), in addition to making the resummed
quantities free of secular divergences, cause the pressure to relax
towards its equilibrium value (one third of the energy density for a
$\phi^4$ theory in four dimensions). Without these fluctuations, the
pressure is simply that of a classical field configuration, and it
oscillates periodically around the equilibrium value, but never
relaxes: therefore there is no one-to-one relationship between the
pressure and the energy density, and thus no equation of state at
LO. The fluctuations of the initial condition for the classical field
in eq.~(\ref{eq:sum1}) produce a decoherence that dampens the
oscillations of the pressure, as illustrated in the figure
\ref{fig:eos}, and after a finite time the pressure converges towards
the value required by the equilibrium equation of state ($3p=\epsilon$).

In \cite{DusliEGV1}, we have also studied the fluctuations of the
energy in a small subsystem. Of course, the total energy in the system
is conserved and therefore does not fluctuate, but its value in a
small subvolume can vary due to exchanges with the surroundings. We
have observed a rapid transition from rather narrow Gaussian
fluctuations at the starting time to broader fluctuations that
are much closer in shape to canonical equilibrium fluctuations at late
times. This transition appears to occurs approximately at the same
time as the relaxation of the pressure.

\subsection{Lattice setup for the study of thermalization}
However, the study performed in \cite{DusliEGV1} does not tell us much
about the microscopic evolution of the system: indeed, since the
energy density and the pressure of the system are intensive quantities
that sum over all the modes of the system, the existence of an
equation of state identical to the equilibrium one does not imply that
the system is in full equilibrium. Equilibrium requires a much more
stringent microscopic arrangement of the system, in which the energy
is distributed among the various modes in a very specific way. In the
present paper, we wish to pursue this study by looking at more
microscopic aspects of the system: namely the existence of
quasi-particle excitations, and the evolution in time of their
momentum distribution. In particular, we would like to know how the
energy of the system, initially introduced in the zero mode via a
spatially homogeneous external source, eventually goes into higher
momentum modes.

A crucial ingredient in this process is the parametric resonance that
exists in the $\phi^4$ scalar field theory, and therefore it is
important that the ultraviolet lattice cutoff be large enough to
comprise the resonance band. If the source $J$ is parametrically
\begin{equation}
J\sim \frac{Q^3}{g}\; ,
\end{equation}
then the resonance band is located at momenta of order $k\sim Q$, and
the UV cutoff $\Lambda$ must therefore obey $Q\lesssim
\Lambda$. Setting up the lattice cutoff in this way is sufficient to
study the evolution of the system at short times, because on these
time scales the occupation number remains small above the resonance
region, as we shall see later in this paper.

This is however not sufficient if we want to study the approach of
the system to thermal equilibrium. In order to see this, recall that
the energy density in the system is parametrically
\begin{equation}
\epsilon \sim \frac{Q^4}{g^2}\; .
\end{equation}
If thermal equilibrium is achieved, this energy density must also be
given by the Stefan-Boltzmann formula (at least for reasonably weak
couplings),
\begin{equation}
\epsilon \sim T^4\; ,
\end{equation}
which tells us that the system would equilibrate at a temperature
\begin{equation}
T\sim \frac{Q}{\sqrt{g}}\; .
\end{equation}
For a numerical simulation to be able to approach the equilibrium
state, the lattice ultraviolet cutoff must be large enough to include
modes of the order of the temperature, which is a more stringent
constraint than simply having the resonance band below the cutoff. At
weak coupling, this implies that the resonance band should be located
towards the soft sector of the lattice spectrum, i.e. in a region
where the lattice mode density is rather low. In order to still have a
significant number of lattice modes inside the resonance band, we used
a larger lattice (of size $20^3$ for the results presented in this
paper), and we have chosen the value of the parameter $Q$ so that the
resonance band is located near $k\approx 1$ (in lattice units, where
the ultraviolet cutoff is at $\Lambda=\sqrt{12}$).

This choice of $Q$ is significantly lower than the value used in
\cite{DusliEGV1} (where we were only interested in the early stages
of the time evolution, dominated by the resonant modes). This means a
lower energy density, and larger time scales. Indeed, since our system
is scale invariant, energy density scales like $Q^4$ and all the times
scale like $1/Q$.  The resulting time evolution of the pressure, for
this choice of $Q$ and a coupling constant $g=1$, is shown in
the figure \ref{fig:eos}.

\section{Spectral function and quasi-particles}
\label{sec:toy:spectral}
\subsection{Definition and leading order}
\label{sec:spectral-function}
Before we study the time evolution of the occupation number in the
system, it is interesting to ask an even more elementary question: can
the system be described in terms of quasi-particles, or on the
contrary does it interact so strongly that no identifiable
quasi-particles show up in its spectrum? To that effect, one can
compute the spectral function, defined as the imaginary part of the
Fourier transform of the retarded propagator\footnote{We assume here
  that the system is spatially homogeneous. This is the case in our
  setup since the source $J$ does not depend on $\x$.}:
\begin{equation}
\rho(\omega,\k;y^0)
\equiv 2\,{\rm Im}\,\int\limits_0^{+\infty} dt d^3\x\; e^{i\omega t}e^{-i\k\cdot\x}\;
G_{_R}(y^0+t,\x,y^0,{\bs 0})\; .
\label{eq:rho-def}
\end{equation}
In this formula, the retarded propagator is normalized so that
\begin{equation}
\Big[\square_x +V^{\prime\prime}(\varphi(x))\Big]\,G_{_R}(x,y)=\delta(x-y)
\end{equation}
for a classical field configuration $\varphi$.  A system in
equilibrium is invariant under translations in time, and therefore its
spectral function defined in this way is in fact independent of the
time $y^0$. However, for transient systems that are not yet in
equilibrium, the spectral function will evolve with time and the $y^0$
dependence is important.

\label{sec:GRLO}
At leading order, we simply need to obtain the retarded propagator in
a classical background field $\varphi(x)$,
\setbox1\hbox to 2cm{
\hfil
\resizebox*{1.85cm}{!}{\includegraphics{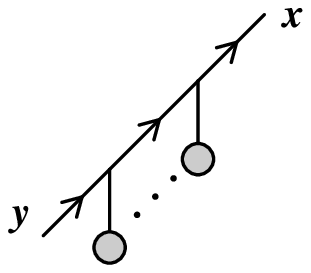}}
\hfil
}
\begin{equation}
G_{_R}^{_{\rm LO}}(x,y)=\sum
\raise -9mm\box1\; ,
\end{equation}
where the grey blobs denote the retarded classical field $\varphi(x)$
and the lines with an arrow the bare retarded propagator (the sum is
over the number of insertions of the classical field, from 0 to
$+\infty$).  In a numerical calculation, the simplest way to compute
this propagator is to consider a small field perturbation $a(x)$ in
that background, that obeys the following equation of motion
\begin{equation}
\Big[\square_x+V^{\prime\prime}(\varphi(x))\Big]\,a(x)=0\; .
\label{eq:eom-fluct}
\end{equation}
The fluctuation $a(x)$ can be related to its value at the time $y^0$
by the following Green's formula
\begin{equation}
a(x)
=\int d^3\y\; \Big[G_{_R}^{_{\rm LO}}(x,y) \left(\partial_{y}^0 a(y)\right)
- \left(\partial_{y}^0G_{_R}^{_{\rm LO}}(x,y)\right)\,a(y)\Big]\; ,
\end{equation}
that involves precisely the propagator we are looking for.  Thus, by
choosing the following initial conditions at time $y^0$,
\begin{equation}
a(y^0,\y)=0\; ,\qquad \partial_y^0 a(y^0,\y)=\delta(\y)\; ,
\label{eq:ic-fluct}
\end{equation}
the perturbation $a(x)$ is precisely the propagator we need in
eq.~(\ref{eq:rho-def}),
\begin{equation}
G_{_R}^{_{\rm LO}}(x^0,\x,y^0,{\bs 0})=a(x^0,\x)\; .
\end{equation}
Thanks to this observation, we reduce the problem of finding the
retarded propagator at leading order to that of solving the equation
(\ref{eq:eom-fluct}) with the initial conditions of
eq.~(\ref{eq:ic-fluct}), which is easily doable
numerically\footnote{On a lattice, the delta function that appears in
  the initial condition for $\partial_y^0 a(y^0,\y)$ becomes a
  Kronecker symbol: the derivative is zero at all points of the
  lattice except at the origin (0,0,0) where it is equal to one.}.

\subsection{Next to leading order and resummation}
\setbox1\hbox to 7.5cm{ \hfil
  \resizebox*{7.35cm}{!}{\includegraphics{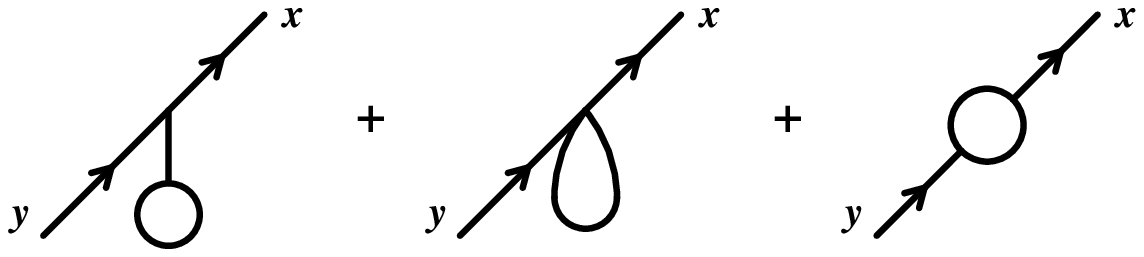}} \hfil } At next to
leading order, three topologies must be evaluated\footnote{Although
  these diagrams seem to involve cubic vertices, this is not
  the case. These vertices where three lines merge are in fact
  proportional to $V^{\prime\prime\prime}(\varphi(x))$, and are thus
  proportional to an extra $\varphi(x)$ that does not appear
  explicitly in the diagrammatic representation.}:
\begin{equation}
G_{_R}^{_{\rm NLO}}(x,y)=
\raise -11mm\box1\; ,
\end{equation}
where the both the propagators and the vertices are dressed by the
classical field $\varphi(x)$. The first topology is the same as the
retarded propagator at leading order, in which one of the $\varphi$
insertions has been replaced by a 1-loop tadpole $\beta$ defined in
eq.~(\ref{eq:fluctuations}). Given that this tadpole is given by (see
the eq.~(39) in \cite{GelisLV3})
\begin{equation}
\beta(x)
=
\Big[
\int d^3\u\; \beta\cdot{\mathbbm T}_\u
+\frac{1}{2}\int d^3\u d^3\v\;\int\frac{d^3\k}{(2\pi)^3 2k}
[a_{+\k}\cdot{\mathbbm T}_\u][a_{-\k}\cdot{\mathbbm T}_\v]
\Big]\,\varphi(x)\; ,
\label{eq:tadpole}
\end{equation}
it is easy to check that this contribution is related to the leading
order one by the following functional identity
\begin{equation}
G_{_R}^{_{\rm NLO1}}(x,y)=
\Big[
\int d^3\u\; \beta\cdot{\mathbbm T}_\u
+\frac{1}{2}\int d^3\u d^3\v\;\int\frac{d^3\k}{(2\pi)^3 2k}
[a_{+\k}\cdot{\mathbbm T}_\u][a_{-\k}\cdot{\mathbbm T}_\v]
\Big]_{{\rm same\ }\varphi}\,G_{_R}^{_{\rm LO}}(x,y)\; ,
\label{eq:GRNLO1}
\end{equation}
where the subscript `same $\varphi$' indicates that the two operators
${\mathbbm T}_\u{\mathbbm T}_\v$ in the second term should act on the
same field $\varphi$, i.e.
\begin{equation}
\Big[{\mathbbm T}_\u{\mathbbm T}_\v\Big]_{{\rm same\ }\varphi} \varphi(x_1)\cdots\varphi(x_n)
\equiv
\sum_{i=i}^n
\varphi(x_1)\cdots\varphi(x_{i-1})
\Bigg[\Big[{\mathbbm T}_\u{\mathbbm T}_\v\Big]\varphi(x_i)\Bigg]
\varphi(x_{i+1})\cdots\varphi(x_n)\; .
\end{equation}
By using the formula
\begin{equation}
G_{+-}^{_{\rm LO}}(x,y)=\int \frac{d^3\k}{(2\pi)^3 2k}\;a_{+\k}(x)a_{-\k}(y)\;,
\label{eq:G+-}
\end{equation}
the second topology can be written as follows,
\begin{equation}
G_{_R}^{_{\rm NLO2}}(x,y)=
\Big[\frac{1}{2}\int d^3\u d^3\v\;\int\frac{d^3\k}{(2\pi)^3 2k}
[a_{+\k}\cdot{\mathbbm T}_\u][a_{-\k}\cdot{\mathbbm T}_\v]
\Big]_{{\rm same\ }V^{\prime\prime}(\varphi)}\,G_{_R}^{_{\rm LO}}(x,y)\; ,
\label{eq:GRNLO2}
\end{equation}
where the subscript `same $V^{\prime\prime}(\varphi)$' indicates that
the two operators ${\mathbbm T}_\u{\mathbbm T}_\v$ should act on the
same compound $V^{\prime\prime}(\varphi)$ --one operator on each field
of $V^{\prime\prime}(\varphi)$.  The third topology can first be
written as
\begin{equation}
G_{_R}^{_{\rm NLO3}}(x,y)=\int d^4w d^4z\; 
G_{_R}^{_{\rm LO}}(x,w)\Sigma^{\rm 1loop}_{_R}(w,z)G_{_R}^{_{\rm LO}}(z,y)\; ,
\label{eq:self1}
\end{equation}
where $\Sigma_{_R}^{\rm 1loop}$ is the 1-loop retarded self-energy,
\begin{equation}
\Sigma_{_R}^{\rm 1loop}(w,z)
=
\Sigma_{++}^{\rm 1loop}(w,z)
-
\Sigma_{+-}^{\rm 1loop}(w,z)\; .
\end{equation}
Next, one can rewrite this self-energy as
\begin{equation}
\Sigma_{_R}^{\rm 1loop}(w,z)
=
\frac{1}{2}V^{\prime\prime\prime}(\varphi(w))V^{\prime\prime\prime}(\varphi(z))
G_{_R}^{_{\rm LO}}(w,z) \Big[G_{+-}^{_{\rm LO}}(w,z)+G_{-+}^{_{\rm LO}}(w,z)\Big]\; ,
\label{eq:self2}
\end{equation}
where the prefactor $1/2$ is the symmetry factor of the loop. By
combining eqs.~(\ref{eq:self1}) and (\ref{eq:self2}) and by using
(\ref{eq:G+-}), one can finally prove
\begin{equation}
G_{_R}^{_{\rm NLO3}}(x,y)=
\Big[
\frac{1}{2}\int d^3\u d^3\v\;\int\frac{d^3\k}{(2\pi)^3 2k}
[a_{+\k}\cdot{\mathbbm T}_\u][a_{-\k}\cdot{\mathbbm T}_\v]
\Big]_{{\rm distinct\ }\varphi's}\,G_{_R}^{_{\rm LO}}(x,y)\; ,
\label{eq:GRNLO3}
\end{equation}
where the qualifier `distinct $\varphi$'s' indicates that the two
operators ${\mathbbm T}_\u{\mathbbm T}_\v$ must act on two fields
$\varphi$'s that are inserted at different points on the LO propagator
$G_{_R}^{_{\rm LO}}$. Adding eqs.~(\ref{eq:GRNLO1}), (\ref{eq:GRNLO2}) and
(\ref{eq:GRNLO3}) therefore simply lifts any restriction on the action
of these operators, and we obtain
\begin{equation}
G_{_R}^{_{\rm NLO}}(x,y)=
\Big[
\int d^3\u\; \beta\cdot{\mathbbm T}_\u
+\frac{1}{2}\int d^3\u d^3\v\;\int\frac{d^3\k}{(2\pi)^3 2k}
[a_{+\k}\cdot{\mathbbm T}_\u][a_{-\k}\cdot{\mathbbm T}_\v]
\Big]\,G_{_R}^{_{\rm LO}}(x,y)\; .
\label{eq:GRNLOfinal}
\end{equation}
This formula is formally identical to the formula we have obtained
previously for the energy-momentum tensor at NLO, and it leads to the
same pathologies due to the presence of secular divergences. Likewise,
the problem can be cured here by performing the same resummation as in
the case of the energy-momentum tensor, that amounts to exponentiating
the quadratic part of the operator in the square brackets in
eq.~(\ref{eq:GRNLOfinal}) (as in eq.~(\ref{eq:sum})):
\begin{equation}
G_{_R}^{_{\rm resummed}}(x,y)\equiv
\exp\Big[
\frac{1}{2}\int d^3\u d^3\v
\int\frac{d^3\k}{(2\pi)^3 2k}\;
[a_{+\k}\cdot{\mathbbm T}_\u][a_{-\k}\cdot{\mathbbm T}_\v]
\Big]\;G_{_R}^{_{\rm LO}}(x,y)\, .
\label{eq:GRresummed}
\end{equation}
This resummation amounts to a functional average over Gaussian
fluctuations of the initial condition of the classical field at
$x^0=0$,
\begin{equation}
G_{_R}^{_{\rm resummed}}=
\int\big[D\alpha(\x)D\dot\alpha(\x)\big]\;F[\alpha,\dot\alpha]
\;G_{_R}^{_{\rm LO}}[\varphi_0+\alpha]\, 
\label{eq:GRresummed1}
\end{equation}
where the Gaussian distribution $F[\alpha,\dot\alpha]$ is defined in
eq.~(\ref{eq:gaussian}).  Therefore, in order to compute the resummed
retarded propagator, we should repeat the procedure outlined in the
section \ref{sec:GRLO} for every classical field $\varphi$ obtained
from an ensemble of initial conditions $\varphi_0+\alpha$, where
$\alpha$ samples the Gaussian distribution $F[\alpha,\dot\alpha]$.

\subsection{Numerical results}
\begin{figure}[htbp]
  \begin{center}
    \resizebox*{10cm}{!}{\rotatebox{-90}{\includegraphics{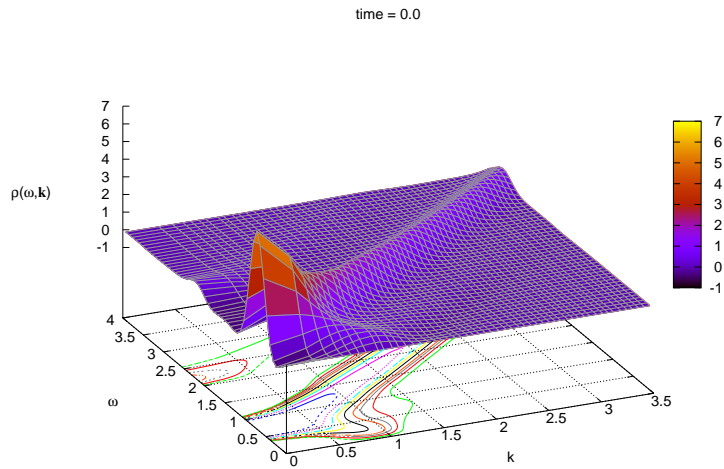}}}
  \end{center}
  \caption{\label{fig:gr0.0}Spectral function
    $\rho(\omega,\k;y^0=0.0)$ at the initial time. The computation is
    done on a $20^3$ lattice, for a coupling constant $g=1$. In this
    plot, $k$ denotes the lattice momentum,
    i.e. $\sqrt{2(3-\cos(2\pi l/L)-\cos(2\pi m/L)-\cos(2\pi n/L))}$ on a $L^3$ lattice
    ($l,m,n$ is the triplet of integers in the range $[0,L-1]$ that
    labels a given momentum state).}
\end{figure}
At the initial time (see the figure \ref{fig:gr0.0}), the spectral
function has a fairly complicated structure. Although the large $\k$
modes have a single spectral peak at $\omega\approx|\k|$, the
situation is richer in the soft sector. There, besides the main branch
that continues to large $\k$, the spectral density exhibits additional
branches. One of them corresponds to a higher mass excitation, and
another one has a mass comparable to the main branch but an anomalous
dispersion such that the frequency decreases while the momentum
increases. Therefore, at early times, the quasi-particle picture is
not a good description of the degrees of freedom in the system.

\begin{figure}[htbp]
  \begin{center}
    \resizebox*{6cm}{!}{\rotatebox{-90}{\includegraphics{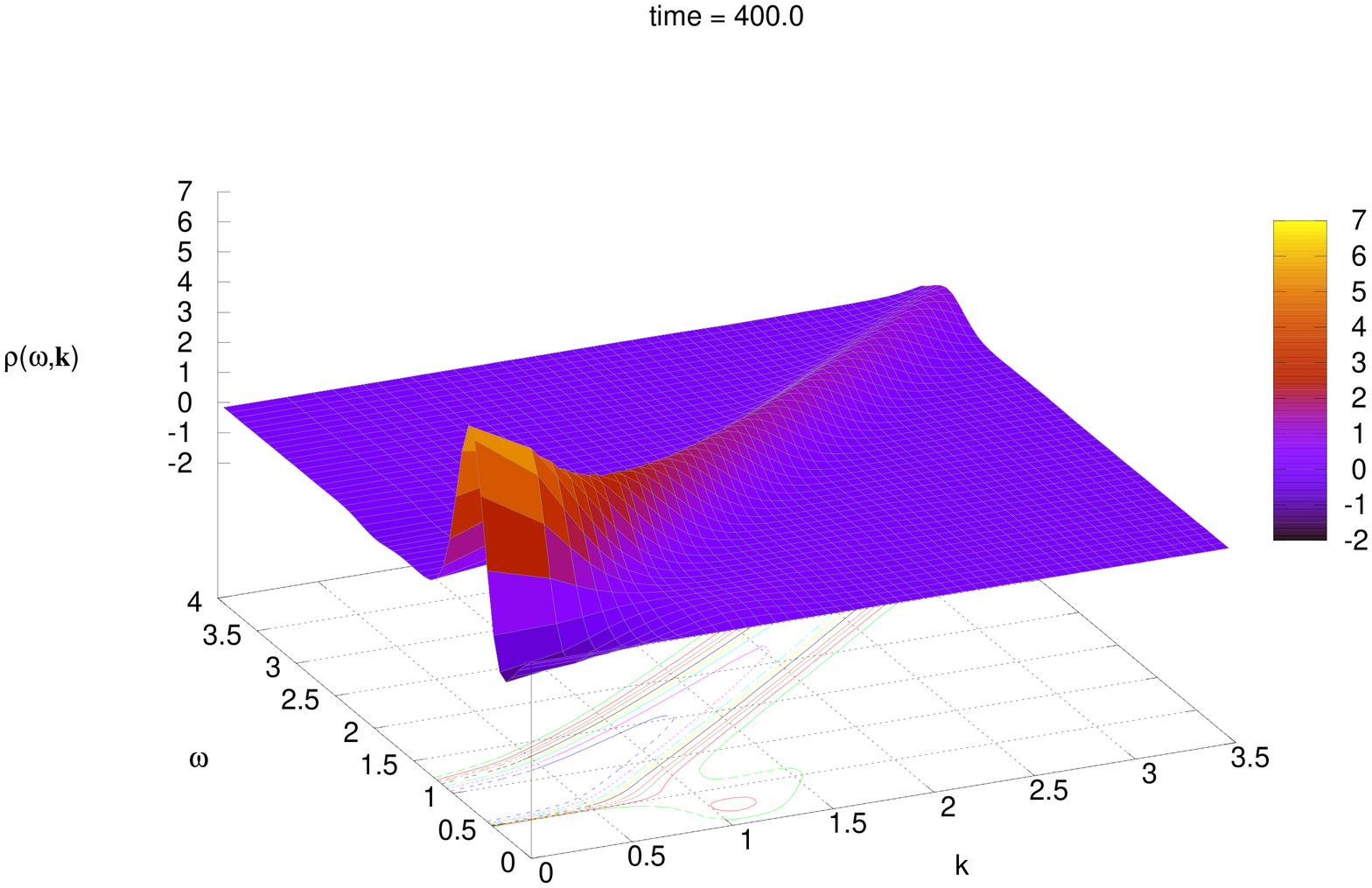}}}
    \hfil
    \resizebox*{6cm}{!}{\rotatebox{-90}{\includegraphics{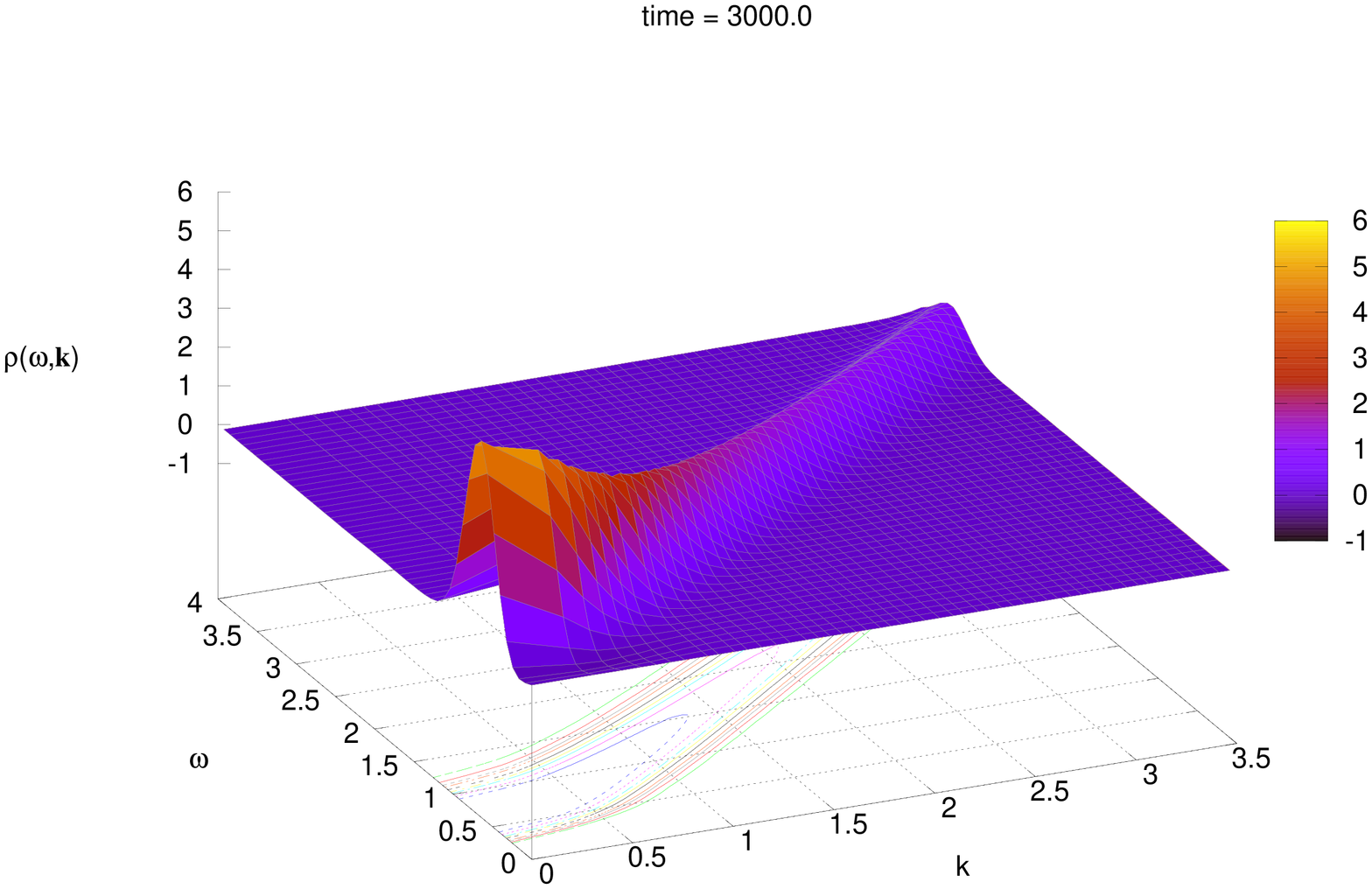}}}
  \end{center}
  \caption{\label{fig:gr-late}Spectral function $\rho(\omega,\k;y^0)$
    at the times $y^0=400$ (left) and $y^0=3000$ (right). The
    computation is done on a $20^3$ lattice, for a coupling constant
    $g=1$.}
\end{figure}
As the time increases, these extra branches in the spectral function
decrease in amplitude and eventually disappear, starting with the
higher mass excitation. Ultimately, only the main excitation remains,
as one can see in the plot on the right of the figure
\ref{fig:gr-late} at a time $y^0=3000$. At intermediate times (such as
$y^0=400$, represented on the plot on the left of the figure
\ref{fig:gr-late}), one gets closer to the spectral function of a
system made of quasi-particles, with only small remnants of the
structures that existed at early times. It is interesting to note that
the characteristic time for the disappearance of the extra branches in
$\rho(\omega,\k;y^0)$ is comparable to the relaxation time of the
pressure, that we have found in the previous section to start at a
time of the order of $y^0\sim 100$.

\subsection{Quasi-particle mass}

\begin{figure}[htbp]
  \begin{center}
    \resizebox*{8cm}{!}{\rotatebox{-90}{\includegraphics{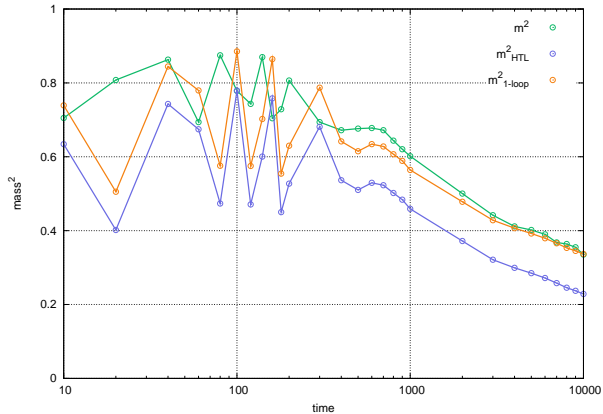}}}
  \end{center}
  \caption{\label{fig:mass}Green line: quasi-particle mass obtained by
    a fit of the main dispersion branch with a function of the form
    $\omega=\sqrt{\k^2+m^2}$. Blue line: 1-loop analytic calculation
    from the occupation number. Red line: 1-loop gap equation that
    resums recursively all the daisy diagrams. (See the text in
    section \ref{sec:qp-mass} for explanations regarding the curves
    labelled $m_{_{\rm HTL}}^2$ and $m_{\rm 1-loop}^2$.)}
\end{figure}

In order to further assess the existence of quasi-particles in the
system, one can try to fit the main branch of the spectral function by
a function of the form $\omega=\sqrt{\k^2+m^2}$. The result of this
fit is shown in the figure \ref{fig:mass}. One sees that the mass $m$
resulting from this fit is not stable until a time $y^0\approx 100$,
and becomes much more regular afterwards. This is in agreement with
the previous qualitative observation that only the main branch of the
spectral function survives after this time. Moreover, after $y^0\ge
1000$, the mass of the quasi-particles that populate the system
decreases slowly with time, indicating that the system is not yet
completely equilibrated (the change in the mass of the quasi-particles
reflects a change in the occupation number of the various modes of the
system, that we will study more directly in the following section).
If one takes as a crude estimate the Hard Thermal Loop
\cite{Pisar2,BraatP1} expression of the medium-generated mass~(see for instance
\cite{KapusG1}, pp 41--45),
\begin{equation}
m^2_{_{\rm HTL}}={g^2}\int\frac{d^3\k}{(2\pi)^3 2 k}\; f_\k\; ,
\label{eq:thermal-mass}
\end{equation}
as a function of the occupation number $f_\k$, we can interpret the
decrease of the mass as a shift of the occupation number from low $\k$
to higher $\k$'s if the total number of quasiparticles ($N\sim\int
d^3\k\,f_\k$) is approximately constant.

A word of caution should be added about the width of the
quasi-particles. The width of the spectral peak in the figures
\ref{fig:gr0.0} and \ref{fig:gr-late} is probably not the physical
width: for practical reasons the numerical computation of the Fourier
transform in time in eq.~(\ref{eq:rho-def}) cannot integrate up to
very large times\footnote{One can check in the free case that this is
  a very singular Fourier transform. At $\k=0$, it is of the form
  $\int_0^{+\infty}dt\;t\;\exp(i\omega
  t)\sim\delta^\prime(\omega)$.}. Thus the width we see in the
resulting plots is to a large extent contaminated by the fact that the
time interval is finite in the numerical calculation (for the physical
width to be visible unambiguously in these plots, the length of the
time interval would have to be much larger than the lifetime of the
quasi-particles).

\section{Occupation number}
\label{sec:occ}
\subsection{Expression in terms of $G_{+-}$ and $G_{-+}$}
Now that we know that at times $x^0\ge 100$, the spectral content of
the system reduces to simple quasi-particles, it makes sense to
compute their occupation number. Recall that the creation and
annihilation operators $a^\dagger_\k,a_\k$ are related to the field
operator $\hat\phi$ via
\begin{eqnarray}
a_\k
&=&
i
\int d^3\x\; e^{ik\cdot x}\; \stackrel{\leftrightarrow}{\partial}_{x^0}\;
\hat\phi(x)
\nonumber\\
a^\dagger_\k
&=&-i
\int d^3\x\; e^{-ik\cdot x}\; \stackrel{\leftrightarrow}{\partial}_{x^0}\;
\hat\phi(x)\; .
\end{eqnarray}
From this, we get the following two reduction formulas
\begin{eqnarray}
\left<a^\dagger_\k a_\k\right>
&=&
\int d^3\x d^3\y\; e^{ik\cdot(x-y)}\;
\stackrel{\leftrightarrow}{\partial}_{x^0}
\stackrel{\leftrightarrow}{\partial}_{y^0}
\;\left.G_{+-}(x,y)\right|_{x^0=y^0}
\nonumber\\
\left<a_\k a^\dagger_\k\right>
&=&
\int d^3\x d^3\y\; e^{ik\cdot(x-y)}\;
\stackrel{\leftrightarrow}{\partial}_{x^0}
\stackrel{\leftrightarrow}{\partial}_{y^0}
\;\left.G_{-+}(x,y)\right|_{x^0=y^0}
\; ,
\end{eqnarray}
with the understanding that the times $x^0$ and $y^0$ are set equal
only after the derivatives have been evaluated.

It turns out to be more straightforward to calculate the sum of these
two expectation values,
\begin{equation}
\left<a^\dagger_\k a_\k+a_\k a^\dagger_\k\right>
=
\int d^3\x d^3\y\; e^{ik\cdot(x-y)}\;
\stackrel{\leftrightarrow}{\partial}_{x^0}
\stackrel{\leftrightarrow}{\partial}_{y^0}
\;\left.G_s(x,y)\right|_{x^0=y^0}
\label{eq:ak-Gs}
\end{equation}
where $G_s\equiv G_{+-}+G_{-+}$, because in our framework the
symmetric propagator $G_s$ is easier to compute than the separate
$G_{\pm\mp}$. The occupation number $f_\k$ is related to the left hand
side of eq.~(\ref{eq:ak-Gs}) by
\begin{equation}
2\omega_\k V (1+2f_\k) = \left<a^\dagger_\k a_\k+a_\k a^\dagger_\k\right>\; ,
\label{eq:comm}
\end{equation}
where $V$ is the volume of the system and $\omega_\k$ the dispersion
relation of the quasi-particles.

\subsection{Calculation of $G_s$}
Let us now see how to compute the symmetric propagator $G_s(x,y)$ at LO,
NLO and in the resummation scheme we have developed to cure the
pathologies related to secular divergences. At leading order, it is
simply given by the product of two classical fields at the points $x$
and $y$,
\begin{equation}
G_s^{_{\rm LO}}(x,y)=2\varphi(x)\varphi(y)\; .
\end{equation}
At next to leading order, $G_s$ is made of two pieces: 
\begin{itemize}
\item[{\bf i.}] a 1-loop correction $\beta$
to one of the factors $\varphi$ of the LO result,
\item[{\bf ii.}] a connected
(tree-level) contribution ${\cal G}_s\equiv{\cal G}_{+-}+{\cal
  G}_{-+}$ that links the points $x$ and $y$,
\end{itemize}
\begin{equation}
G_s^{_{\rm NLO}}(x,y)
=
2\Big[\beta(x)\varphi(y)+\varphi(x)\beta(y)\Big]
+{\cal G}_s(x,y)\; .
\end{equation}
The second term is given by
\begin{equation}
{\cal G}_s(x,y)
=
\int\frac{d^3\k}{(2\pi)^3 2 k}\;
\Big[
a_{+\k}(x)a_{-\k}(y)+a_{-\k}(x)a_{+\k}(y)
\Big]\; .
\end{equation}
The $a_{\pm\k}$'s can be formally related to the classical field
$\varphi(x)$ by (see eq.~(\ref{eq:shift}))
\begin{equation}
a_{\pm\k}(x) =
\int d^3\u\; [a_{\pm\k}\cdot{\mathbbm T}_\u]\;\varphi(x)\; ,
\end{equation}
while for the tadpole $\beta$ we can use eq.~(\ref{eq:tadpole}).
Then, it is straightforward to combine the two terms to obtain
\begin{equation}
G_s^{_{\rm NLO}}(x,y)=
\Big[
\int d^3\u\; \beta\cdot{\mathbbm T}_\u
+\frac{1}{2}\int d^3\u d^3\v\;\int\frac{d^3\k}{(2\pi)^3 2k}
[a_{+\k}\cdot{\mathbbm T}_\u][a_{-\k}\cdot{\mathbbm T}_\v]
\Big]\,G_s^{_{\rm LO}}(x,y)\; .
\label{eq:GsNLOfinal}
\end{equation}
From here, it is clear that one can perform the same resummation,
where one exponentiates the quadratic part of the operator in the
square brackets. This amounts to an average over Gaussian fluctuations
of the initial classical field at $x^0=0$,
\begin{equation}
G_s^{\rm resummed}(x,y)
=
2
\int\big[D\alpha D\dot\alpha]\;F[\alpha,\dot\alpha]\;
\Big[\varphi(x)\varphi(y)\Big]_{\varphi_0+\alpha},
\end{equation}
where the subscript $\varphi_0+\alpha$ indicates the initial condition
used at $x^0=0$ to start the evolution of the classical field
$\varphi$. $F[\alpha,\dot\alpha]$ is the Gaussian distribution of
fluctuations defined in eqs.~(\ref{eq:sum1}) and (\ref{eq:gaussian}).
 
\subsection{Time evolution of $f_\k$}
By combining the previous results, the occupation number obtained in
this resummation scheme can be written as
\begin{equation}
f_\k
=
-\frac{1}{2}+
\frac{1}{2\omega_\k V}
\int\big[D\alpha D\dot\alpha]\;F[\alpha,\dot\alpha]\;
\left|
\int d^3\x\; e^{i\k\cdot\x}\;(\dot\varphi(x^0,\x)+i\omega_\k \varphi(x^0,\x))
\right|^2_{\varphi_0+\alpha}\; .
\label{eq:fk}
\end{equation}
In the evaluation of this formula, we use for the energy
$\omega_\k=\sqrt{\k^2+m^2}$ with the mass fitted in the previous
section (thus, we use a different mass at each time $x^0$).
\begin{figure}[htbp]
  \begin{center}
    \resizebox*{8cm}{!}{\rotatebox{-90}{\includegraphics{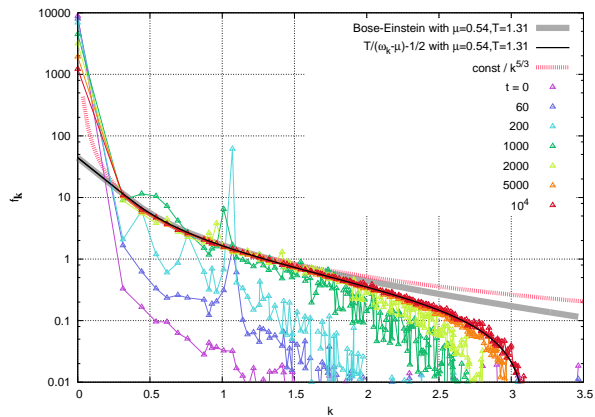}}}
  \end{center}
  \caption{\label{fig:fk}Occupation number $f_\k$ at various times in
    the evolution of the system. The grey band represents a fit by a
    Bose-Einstein distribution. The dashed red band is a fit by a pure
    power law $k^{-5/3}$. The thin black line is a fit by a
    distribution of the form given in eq.~(\ref{eq:class}).}
\end{figure}
The result of this calculation is displayed in the figure
\ref{fig:fk}, where we show the occupation number at various stages of
the time evolution, as well as three fits that we shall discuss shortly.

Let us first briefly describe the main stages of the time
evolution. At the initial time $t=0$, only the zero mode is occupied
and the higher modes have a negligible occupation number. This is a
direct consequence of our setup, where the classical field is
initially driven by a spatially homogeneous source.  Then, shortly
afterwards (this is already visible in the spectrum at $t=20$) one
sees an increase of the occupation in the non-zero modes, concomitant
with a decrease of the occupation in the zero mode (barely visible in
the figure, due to the logarithmic vertical scale). The increase of
the non-zero modes is most pronounced in a narrow band of $k$, where
it peaks more than an order of magnitude above the rest of the
curve. One can check\footnote{See the appendix B of \cite{DusliEGV1}.}
that this band of $k$ coincides with the band of parametric resonance
that we have discussed in detail in \cite{DusliEGV1}. Thus, it appears
that the dominant physics at early times is that of resonance, which
leads to a quick increase of the occupation number in a narrow region
of $\k$. After $t=1000$, the resonance peak has disappeared and the
evolution becomes fairly slow.

Let us now discuss fits of the occupation number, that are represented
in the figure \ref{fig:fk}. The first two are a fit by a Bose-Einstein
distribution,
\begin{equation}
f_{_{\rm BE}}(k)
=\frac{1}{e^{\beta(\omega_\k-\mu)}-1}\; ,
\label{eq:BEmu}
\end{equation}
and a fit by a classical distribution of the form
\begin{equation}
f_{\rm class}(k)=\frac{T}{\omega_\k-\mu}-\frac{1}{2}\;. 
\label{eq:class}
\end{equation}
Interestingly, the best fit we could achieve with a Bose-Einstein
distribution required a non-zero chemical potential. Although the
particle number has no reason to be conserved in this theory (there is
no symmetry protecting it), this suggests that changes of the particle
number are slow compared to the evolution of the distribution in
momentum space: a chemical potential at the latest times we have
considered indicates that the particle number has not yet reached its
equilibrium value (and its positive sign means that we have a particle
excess). 
\begin{figure}[htbp]
  \begin{center}
    \resizebox*{8cm}{!}{\rotatebox{-90}{\includegraphics{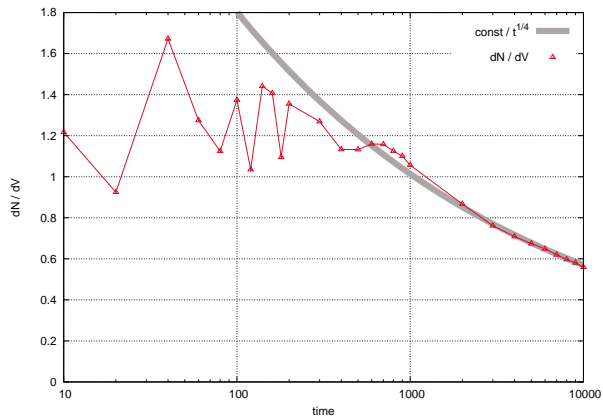}}}
  \end{center}
  \caption{\label{fig:density}Time evolution of the quasi-particle
    density in the system. Gray band: fit of the tail with a power law
    $t^{-1/4}$.}
\end{figure}
At weak coupling, this is rather natural: inelastic processes have a
much smaller rate than the elastic ones\footnote{For the $\phi^4$
  scalar theory that we consider here, $\sigma_{\rm el}\sim g^4$,
  while $\sigma_{\rm inel}\sim g^8$. The hierarchy between the elastic
  and inelastic time-scales is certainly less pronounced in QCD (see
  \cite{ArnolDM1}) and there it is unclear whether there is enough
  time for the formation of a transient state that has a non-zero
  chemical potential.}, and therefore at intermediate time scales the
number of particles is an approximately conserved quantity.  In order
to check this hypothesis, we can evaluate the number density by
summing the occupation number over all the modes $\k$. This has been
done in the figure \ref{fig:density}.  One sees indeed that, after a
period of somewhat erratic evolution (that roughly corresponds to the
time necessary to have well defined quasi-particles in the system),
the number density decreases very slowly at late times, as a small
negative power of time.

\begin{figure}[htbp]
  \begin{center}
    \resizebox*{8cm}{!}{\rotatebox{-90}{\includegraphics{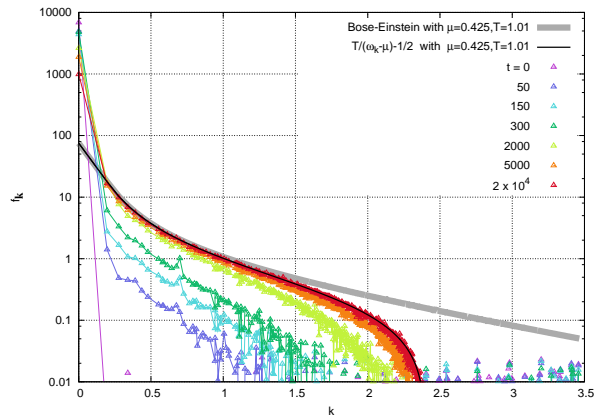}}}
  \end{center}
  \caption{\label{fig:fk_small}Occupation number $f_\k$ at various
    times in the evolution of the system, for a system initialized
    with a smaller energy density.}
\end{figure}

It is also obvious from the figure \ref{fig:fk} that a Bose-Einstein
distribution does not fit well the occupation number in the tail at
large $k$. In fact, the contrary would have been surprising, since our
computation is essentially semi-classical. Naively, one may expect to
obtain a classical distribution of the form $T/(\omega_\k-\mu)$
(again, a non-zero $\mu$ is allowed if number changing processes are
very slow), but one can check that such a distribution does not
produce a better fit of the tail. At first sight, one could be tempted
to blame this drop in the tail on the rarefaction of the lattice modes
at large $k$ (see for instance the figure 15 in
\cite{DusliEGV1}). However, this assumption does not hold if one does
the same simulation with lower physical scales, as is done in the
figure \ref{fig:fk_small}. On this figure, one sees the same drop, now
occurring at a smaller value of $k$. If the drop was caused by lattice
artifacts, one would expect it to occur at a fixed value of $k$ (in
lattice units), no matter what the physical scales are.

It turns out that this drop has a rather trivial explanation. Firstly,
note that a very good fit is obtained with the distribution given in
eq.~(\ref{eq:class}) (the thin black line in the figures \ref{fig:fk}
and \ref{fig:fk_small}), that differs from the naive classical
distribution by an extra $-1/2$ term. This extra term, that makes the
fit considerably better, has a simple origin: it comes from the
$-1/2$ in the equation (\ref{eq:fk}), which in the derivation of the
formula for $f_\k$ can be traced back to the non-zero commutator
between creation and annihilation operators. Keeping this $-1/2$
correction in the definition of the occupation number for a
semi-classical calculation is to a large extent an arbitrary
choice. Indeed, such an approximation is expected to reproduce
correctly the underlying quantum theory only in the region where the
occupation number is sufficiently large. When this is the case, one
has $f_\k+1/2\approx f_\k$, and therefore this $1/2$ is not very
significant. This also means that the drop of the occupation number at
$k\ge 2$ in the figure \ref{fig:fk}, although perfectly understandable
in our semi-classical approximation, is obviously not a physical
feature of the underlying quantum theory\footnote{Interestingly
  however, the $-1/2$ term in eq.~(\ref{eq:class}) is nothing but the
  second term in the expansion of the Bose-Einstein distribution in
  powers of $(\omega_\k-\mu)/T$. Therefore, at a formal level, keeping
  this $1/2$ correction in the present semi-classical computation
  gives a better approximation of the full quantum theory.  This point
  was already discussed extensively in \cite{MuellS1,Jeon3} in the
  context of the Boltzmann equation.}. In our setup, there is one
advantage in keeping the $-1/2$ in eq.~(\ref{eq:fk}) though: if one
does the same computation with a vanishing source $J=0$, one gets
identically $f_\k=0$, which is of course the exact answer. Without
this $-1/2$, one would have obtained $f_\k=1/2$.

\subsection{Kolmogorov turbulence}
At $t\approx 200$, the modes in the resonance band reach their maximal
occupancy, and start to subside afterwards, while the other non-zero
modes continue to increase. While the resonance peak progressively
disappears, one sees in the figure \ref{fig:fk} that the occupation
curves tend to accumulate in the intermediate $k$ range on a fixed
line that is well fitted by a power law $k^{-5/3}$. In this regime,
the zero mode continues to decrease, while the occupation curve
extends slowly into the hard region.

Such a scaling with an exponent $-5/3$ in the power law is well know
in the physics of turbulence (see the first part of \cite{Berna1} for
instance). Typically, in Kolmogorov's turbulence, the energy cascades
to the hard modes from a source localized in the soft sector, with an
intermediate stationary distribution in between, that follows a power
law $k^{-5/3}$. In our case, the zero mode plays the role of this
source, since it was initially the only occupied mode. In contrast to
the usual setup in the study of Kolmogorov's turbulence \cite{Berna1},
our system is closed and eventually the zero mode will run out of
energy and will not be able to feed the cascade anymore. However, in
our simulation, we have not reached the time at which this starts to
happen.

\subsection{Bose-Einstein condensation}
In the figure \ref{fig:fk}, we saw that the occupation number at late
times is best fitted by a distribution of the form of
eq.~(\ref{eq:class}). This fit however calls for two comments:
\begin{itemize}
\item[{\bf i.}] the occupation number of the zero mode lies above the
  curve provided by this fit,
\item[{\bf ii.}] the best value of the chemical potential ($\mu=0.54$)
  is close to the mass of the quasi-particles at this time, $m=0.58$.
\end{itemize}
These two seemingly unrelated facts have  a common
interpretation. As we have seen before, a chemical potential arises
because the number of quasi-particles evolves very slowly in this
system, and a positive $\mu$ is the reaction of the system to
accommodate an excess of particles. However, it is clear from
eqs.~(\ref{eq:BEmu}) and (\ref{eq:class}) that $\mu$ cannot be larger
than the mass $m$ -- otherwise, the occupation number would become
negative near $k=0$. But having an upper bound on the chemical
potential implies an upper bound on the particle density that these
distributions can describe. What if the particle excess in the system
is so large that the density is larger than this upper bound?  When
this happens, the excess of particles condenses on the zero mode, a
phenomenon known as Bose-Einstein condensation. Dynamically, the
system evolves towards a distribution made of two
components\footnote{Here we have written the quantum version of the
  distribution, but it has an analogue in our semi-classical
  approximation, where the first term is replaced by $f_{\rm
    class}(\k)$. This is discussed in more detail in the appendix
  \ref{sec:classical}.},
\begin{equation}
f_\k =  \frac{1}{e^{\beta(\omega_\k-m)}-1}+f_0\,\delta(\k)\; ,
\label{eq:BEmu+cond}
\end{equation}
i.e. the chemical potential settles to the maximal value $\mu=m$, and
the extra particles go into the zero mode\footnote{Using the Boltzmann
  equation, it is easy to see that $\k=0$ is the only mode where the
  extra particles can go. If the particles in excess occupy non-zero
  modes, then one does not have a fixed point of the Boltzmann
  equation.}  $\k=0$. As one can see from the fit of the figure
\ref{fig:fk}, the occupation number at low $\k$ appears to be
precisely of the form of eq.~(\ref{eq:BEmu+cond}).

\begin{figure}[htbp]
  \begin{center}
    \resizebox*{5.5cm}{!}{\rotatebox{-90}{\includegraphics{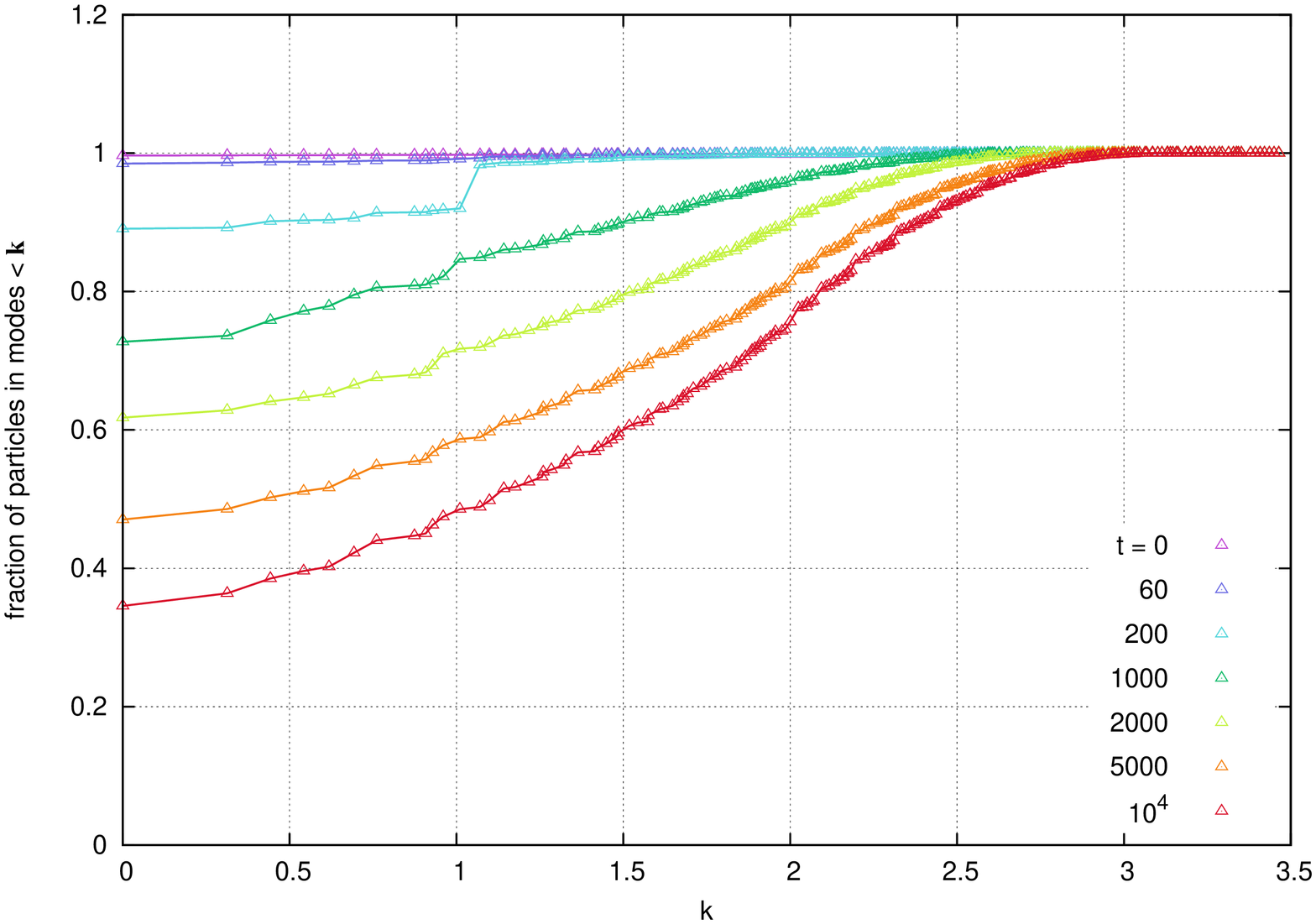}}}
    \hfil
    \resizebox*{5.5cm}{!}{\rotatebox{-90}{\includegraphics{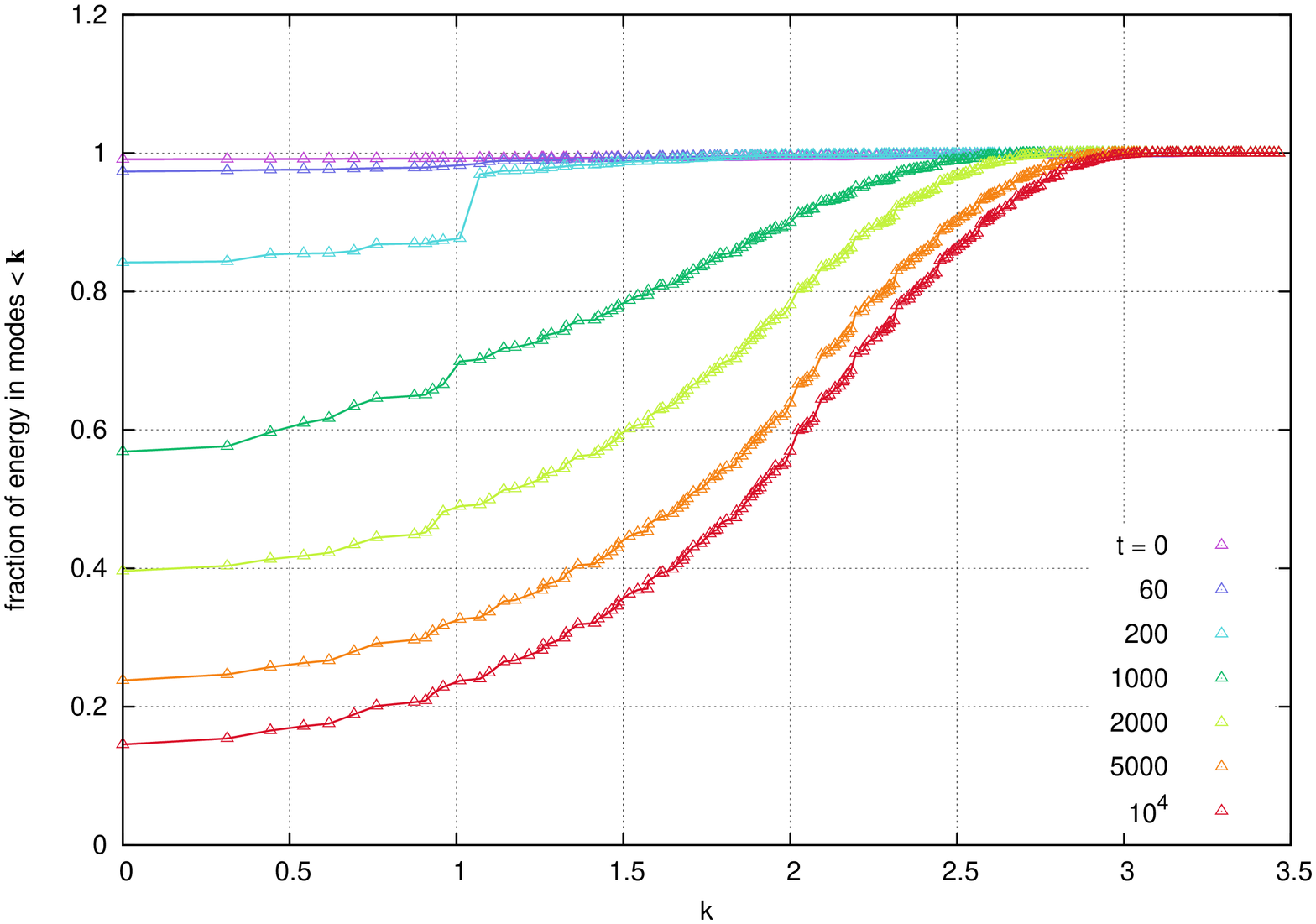}}}
  \end{center}
  \caption{\label{fig:efrac}Left: fraction of particles contained in
    the modes $|\l|\le |\k|$, at various stages of the time
    evolution. Right: fraction of energy contained in the modes
    $|\l|\le |\k|$.}
\end{figure}

From the knowledge of the occupation number, it is easy to compute
what fraction of the number of particles and what fraction of the
total energy are contained in the zero mode at various times. This
information is provided in the two plots of the figure
\ref{fig:efrac}.
At early times (up to $t\sim 100$), all the particles and all the
energy are stored in the zero mode, as a consequence of our initial
condition. At intermediate times (e.g. at $t=200$), a large fraction
of the energy is still in the zero mode, and the remainder is almost
entirely in the resonance band.  At the latest time we have considered
($x^0=10^4$ lattice units), the zero mode still contains about $35\%$
of the particles and $15\%$ of the energy.

\begin{figure}[htbp]
  \begin{center}
    \resizebox*{8cm}{!}{\rotatebox{-90}{\includegraphics{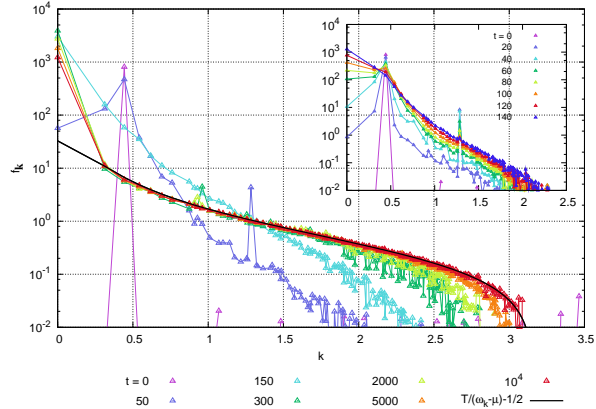}}}
  \end{center}
  \caption{\label{fig:fk_cond_k1}Occupation number $f_\k$ at various
    times, for a system initialized in the modes
    $(k_x,k_y,k_z)=(1,1,0)$ and $(-1,-1,0)$. In the top right inset,
    we show the behavior at short times.}
\end{figure}

On could argue that our computation does not demonstrate Bose-Einstein
condensation, because we started from an initial condition in which
all the energy is already stored in the zero mode. What if we had
started from a situation where the zero mode is empty? We have done
that in the figure \ref{fig:fk_cond_k1}, in which the energy of the
system is initially contained in the modes\footnote{We initialize the
  system in two opposite modes so that there is no net momentum in the
  system.}  $(k_x,k_y,k_z)=(1,1,0)$ and $(-1,-1,0)$ (the total energy
being exactly the same as in the figure \ref{fig:fk}).
Thus, at the initial time, the occupation number has a delta peak at a
single, non-zero, energy. But then, one sees that a non-zero
occupation number develops in the zero mode (and in other modes as
well), to reach very large values in a rather short time. After this
rapid transient regime, all trace of the original peak has been washed
out. At late times, the distribution has become identical to the one
encountered in the figure \ref{fig:fk}: all the modes except $\k=0$
are described by a function of the form of eq.~(\ref{eq:class}),
and there is a particle excess in the zero mode. This study strongly
suggests that Bose-Einstein condensation indeed occurs in this
system\footnote{Of course, since the study is performed on a lattice,
  that has by definition a discrete spectrum, it is impossible to tell
  whether the distribution has a true $\delta(\k)$ term at the origin
  or whether it is a strongly peaked but otherwise regular function.},
when it is initially over-occupied.

\section{Further tests of the quasi-particle description}
\label{sec:qp-mass}
\subsection{Quasi-particle mass}
From the occupation number, we can further test the quasi-particle
description of the system. A simple check is to compute the
medium-induced mass of the quasi-particles, assuming that perturbation
theory applies. The Hard Thermal Loop contribution to this mass has
already been given in eq.~(\ref{eq:thermal-mass}), and we have
represented this quantity in the figure \ref{fig:mass} (blue curve),
along with the mass obtained by fitting the location of the peak in
the spectral function. In the region where quasi-particles are well
defined ($y^0\ge 100$), we see that the HTL value of the mass is
systematically lower than the observed one. We can improve this
result by including also the 1-loop vacuum contribution to the
mass. At 1-loop in a $\phi^4$ theory, this is given by a tadpole graph
whose expression can be written as
\begin{equation}
m^2_{\rm 1-loop} = g^2\int \frac{d^3\k}{(2\pi)^32k}\;
\Big(\frac{1}{2}+f_\k\Big)\; .
\end{equation}
Including the vacuum contribution to the mass improves significantly
(see the red curve in the figure \ref{fig:mass}) the agreement between
the theoretical prediction and the fit, indicating that higher-order
corrections are presumably rather small for this value of the coupling
($g=1$). Thus, it appears that the quasi-particle description is quite
consistent: indeed, the occupation number computed from the fields
themselves, when inserted into the 1-loop formula for the effective
mass, reproduces very well the mass measured by fitting the peak in
the spectral function.

\subsection{Residual interaction energy}
A quasi-particle description of a system is useful only if the
residual interactions between the quasi-particles are weak -- in other
words, if the main effect of the interactions is simply to alter the
properties of the particles (e.g. by generating an effective mass).
\begin{figure}[htbp]
  \begin{center}
    \resizebox*{8cm}{!}{\rotatebox{-90}{\includegraphics{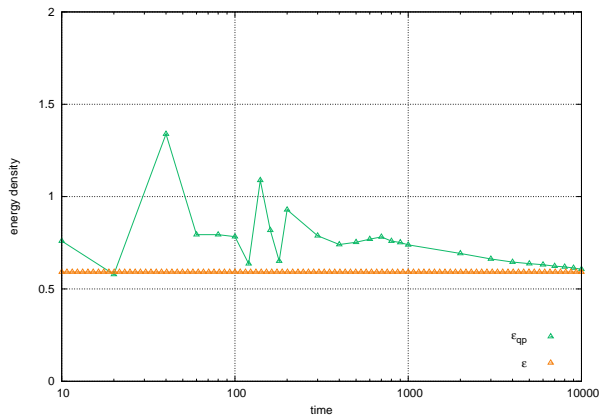}}}
  \end{center}
  \caption{\label{fig:energy-1}Comparison between the actual energy
    density of the system ($\epsilon$) and the energy carried by the
    quasi-particles ($\epsilon_{\rm qp}$) if one neglects their
    interactions.}
\end{figure}
This can be tested by computing the energy density of the system by
summing the energies of its quasi-particles, i.e. by assuming that
they have no residual interactions\footnote{In our framework,
  replacing the true energy density by $\epsilon_{\rm qp}$ is
  equivalent to substituting the expectation value of the interaction
  energy $\left<V(\varphi)\right>$ by
  $\frac{1}{2}m^2\left<\varphi^2\right>$, i.e. to a mean-field
  approximation.},
\begin{equation}
\epsilon_{\rm qp}
\equiv
\int\frac{d^3\k}{(2\pi)^3}\; f_\k\; \sqrt{\k^2+m^2}\; .
\end{equation}
By comparing this quasi-particle energy to the actual energy density,
$\epsilon\equiv \big<T^{00}\big>$, we can estimate the interaction
energy of the quasi-particles and therefore the strength of their
residual interactions. The result of this comparison is shown in the
figure \ref{fig:energy-1}. We see that the true energy of the system
is always below the energy of its quasi-particles, indicating that the
residual interactions are attractive -- which is indeed a standard
result of a $\phi^4$ field theory. Moreover, as the time increases,
the energy of the quasi-particles gets closer to the true energy,
meaning that the quasi-particle description is better at late times.

\subsection{Entropy production}
\begin{figure}[htbp]
  \begin{center}
    \resizebox*{8cm}{!}{\rotatebox{-90}{\includegraphics{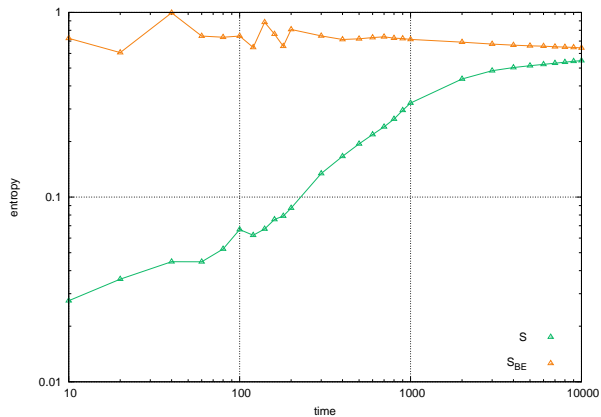}}}
  \end{center}
  \caption{\label{fig:entropy-1}Green curve $(S)$: time evolution of
    the entropy density $s$ defined in eq.~(\ref{eq:entropy}). Red
    curve $(S_{_{\rm BE}})$: entropy of a non-interacting gas of
    bosons of equal energy density, in thermal equilibrium.}
\end{figure}

From the occupation number, it is also possible to compute the entropy
density,
\begin{equation}
s\equiv
\int\frac{d^3\k}{(2\pi)^3}\; \Big[(1+f_\k)\ln(1+f_\k)-f_\k\ln f_\k\Big]\; .
\label{eq:entropy}
\end{equation}
The time evolution of this quantity is shown in the figure
\ref{fig:entropy-1} (green curve). One sees that the entropy density
is multiplied roughly by a factor 20 during the evolution of the
system (the initial value is low in our setup because the occupancy is
entirely localized in the zero mode at $t=0$). In the red curve, we
have displayed the entropy that would have a gas of free bosons of
equal energy density at thermal equilibrium\footnote{Its small
  variations with time are due to the fact that the mass of the
  quasi-particles is not constant.}. The true entropy of the system
gets close to the equilibrium entropy, but not exactly equal even at
the largest times we have considered (the discrepancy remains of the
order of $10-20\%$). This difference is presumably a combination of
two factors: (i) the fact that the system is not yet fully
equilibrated, and (ii) the residual interactions of its
quasi-particles.

The increase of the entropy defined by eq.~(\ref{eq:entropy}) may seem
paradoxical at first sight, given the way it has been obtained in our
framework. Indeed, at the microscopic level, our system is described
as an ensemble of classical field configurations that evolve according
to the Euler-Lagrange equation of motion. This equation of motion is
invariant under time reversal $t\to -t$, yet the entropy $s$ computed
in this system is clearly not invariant under this transformation --
despite the fact that it is a functional of classical fields that have
a time-reversible evolution. Let us recall here that the ensemble of
classical field configurations that describe the system in our
framework evolves according to the Liouville equation,
\begin{equation}
\partial_t {\cal F}_t+\{{\cal F}_t,H\}=0\; .
\label{eq:Liouville-1}
\end{equation}
Instead of eq.~(\ref{eq:entropy}), one could have defined an entropy
based on the probability distribution ${\cal
  F}_t[\varphi,\dot\varphi]$ of the field configurations in
phase-space,
\begin{equation}
{\cal S}\equiv 
-\int [D\varphi D\dot\varphi]\; {\cal F}_t[\varphi,\dot\varphi]\;
\ln{\cal F}_t[\varphi,\dot\varphi]\; ;
\label{eq:entropy-1}
\end{equation}
and it is easy to see that it is constant in time thanks to
Liouville's theorem (see the appendix \ref{app:liouville}). This means
that, if one were able to determine the field configuration of the
system at a given time, there would be no entropy increase because
everything would be known about the microscopic state of the
system. In contrast, the definition (\ref{eq:entropy}) is the
appropriate definition when one knows only the single particle
distribution in the system. Compared to eq.~(\ref{eq:entropy-1}), a
lot of information about the microscopic state of the system has been
discarded by doing this. This coarse graining is the reason why the
entropy given by (\ref{eq:entropy}) increases with time, while at the
microscopic level the field configurations evolve via time reversible
equations.

\section{Summary and outlook}
\label{sec:t-evol}
In this paper, we have pursued the study started in \cite{DusliEGV1}
of the time evolution of a system of scalar fields in a fixed volume
box, in a resummation scheme that has been devised to cure the problem
of secular divergences caused by parametric resonance. Although it is
a resummation of quantum corrections that arise via loops, we have
shown that this resummation can also be formulated as an average over
classical field trajectories, with a Gaussian ensemble of initial
conditions. This latter formulation is the one we adopt for a
practical numerical implementation. In \cite{DusliEGV1}, we have
focused mostly on the evolution of the energy-momentum tensor of the
system, and we have shown that the pressure relaxes towards its equilibrium
value, thanks to a decoherence mechanism due to the fluctuating
initial conditions.

In the present paper, we have studied more microscopic properties of
the system, in order to understand in more detail its time
evolution. In particular, one question that was left unanswered in
\cite{DusliEGV1} is whether the system is in local thermal
equilibrium at the time its pressure tensor reaches the equilibrium
form. First of all, we have studied the existence of quasi-particles in the
spectrum of the theory, by computing the spectral function of the
system. Next, after having identified quasi-particle modes, we have
computed their occupation number and the associated entropy.

\begin{figure}[p]
  \begin{center}
    \resizebox*{13.0cm}{!}{\rotatebox{-90}{\includegraphics{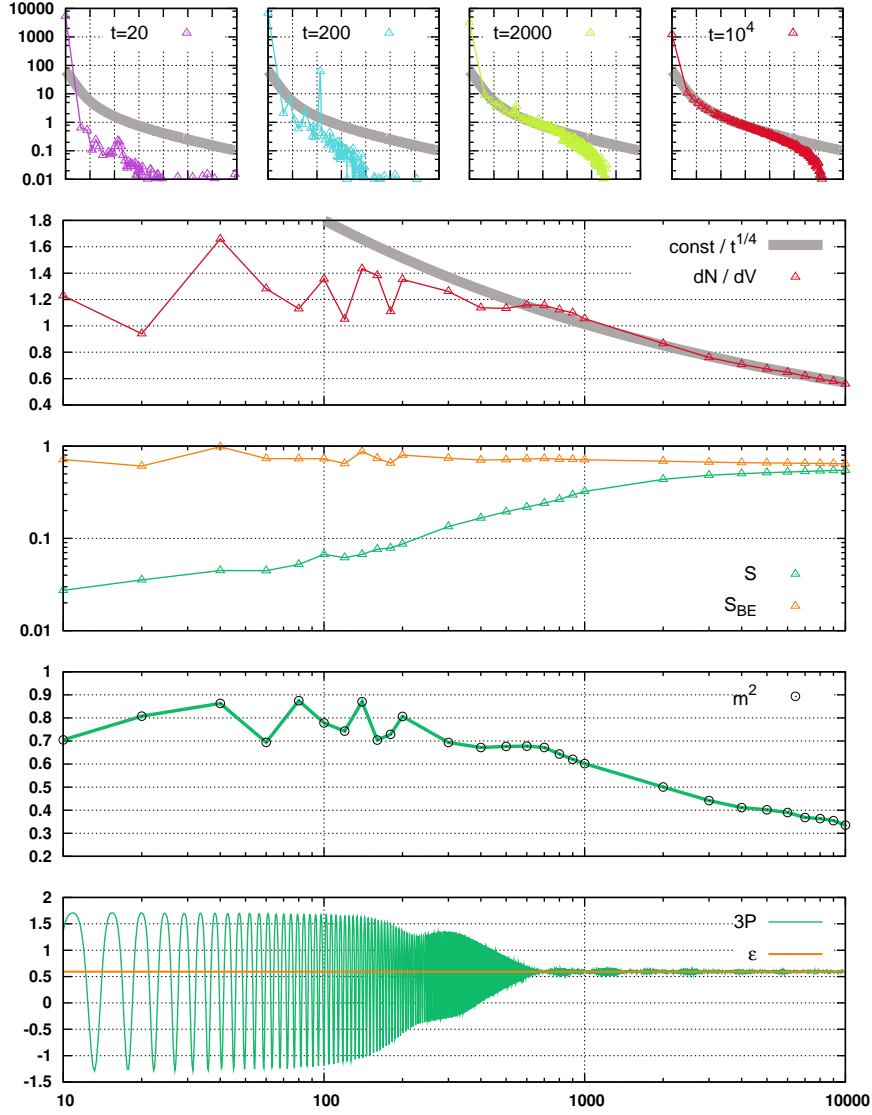}}}
  \end{center}
  \caption{\label{fig:summary}(Panels numbered 1 to 5 from the bottom
    to the top.) Panel 1: time evolution of the pressure. Panel 2:
    time evolution of the quasi-particle mass.  Panel 3: time
    evolution of the entropy, compared to the entropy of a gas of same
    energy density at thermal equilibrium. Panel 4: time evolution of
    the quasi-particle density. Panel 5: occupation number at various
    stages of the time evolution (the gray band is a fit at the latest
    time by a Bose-Einstein distribution with a chemical potential).}
\end{figure}

Let us first summarize the main results of this paper in a synthetic
way, by displaying in parallel the time evolutions of the various
quantities that we have considered separately so far. This is shown in
the figure \ref{fig:summary}.
From these plots, it appears that one can divide the time evolution in
three stages that are qualitatively distinct\footnote{The numerical
  values of the times quoted here are not absolute, but depend on the
  energy density in the system. Indeed, in a scale invariant theory,
  all the time-scales vary like $\epsilon^{-1/4}$. Moreover, these
  time-scales depend on the coupling constant $g^2$, and decrease as
  the coupling increases.}:
\begin{itemize}
\item[{\bf i.}]$0\le t\le 100$~:~ at these early times, the pressure
  of the system has not yet started to relax towards its equilibrium
  value $p=\epsilon/3$. Moreover, quasi-particles are not a good
  description of the system (their mass is not well defined, and there
  are extra spurious branches in the spectral function). The entropy
  displays only a very moderate growth during this era, the occupation
  number starts rising in the resonance band almost immediately, but
  the energy is still almost entirely contained in the zero mode,
\item[{\bf ii.}] $100\le t\le 600$~:~ this intermediate period starts
  when the occupancy in the resonance band reaches its maximum and
  starts to subside, while the occupation number rises in the other
  momentum modes. During this stage, the zero mode still contains the
  largest share of the total energy, and the remainder is contained
  predominantly in the resonant modes.  The mains features of this era
  are the relaxation of the pressure towards its equilibrium value,
  and an important growth of the entropy. In this era, there are well
  defined quasi-particles, with a mass that is almost constant in
  time,
\item[{\bf iii. }] $600\le t$~:~ in the late stages of the time
  evolution, the pressure is the equilibrium one, and the entropy
  displays only a marginal growth. However, the system is not yet
  fully equilibrated: the mass of the quasi-particles shows a clear
  decrease, while the occupation number continues to slowly expand
  towards higher momenta. During this stage, the occupation number
  shows some signs of Kolmogorov scaling, but perhaps even more
  compelling is the fact that it seems to evolve as would a system
  dominated by elastic collisions and which is overpopulated compared
  to equilibrium -- i.e. by developing a chemical potential equal to
  the mass and a Bose condensate at zero momentum. At the same time,
  the energy initially contained in the zero mode is progressively
  distributed among the higher modes.
\end{itemize}

A very interesting observation is the appearance of a chemical
potential, that we interpreted as resulting from a particle excess
(compared to the value the particle density should have in
equilibrium) combined to a fairly slow rate of inelastic
processes. Eventually, the inelastic processes will wipe out the
initial particle excess. However, because they are slow, there is an
extended regime where the occupation number settles on a form that has
a chemical potential. Moreover, for the bosonic statistics, the
chemical potential cannot be larger than the mass of the
quasi-particles. This implies that if the particle excess is too
large, it cannot be accommodated solely by a chemical potential and
the system stabilizes itself by the formation of a condensate at zero
momentum. Our numerical results confirm this, and moreover indicate
that this condensation occurs very early in the time
evolution. Interestingly, the condition of overpopulation is also
realized for the gluons produced initially in heavy ion
collisions. Indeed, at a time $Q_s^{-1}$ (where $Q_s$ is the
saturation momentum), the energy density is $\epsilon\sim Q_s^4/g^2$
and the gluon number density is $n\sim Q_s^3/g^2$. Thus one has the
dimensionless ratio $n\epsilon^{-3/4}\sim g^{-1/2}\gg 1$, that should
be of order 1 in chemical equilibrium. Depending on the strength of
the number-changing processes, one may also expect the (transient)
appearance of a gluonic chemical potential and the formation of a
gluon condensate at zero momentum.

One of the virtues of the formalism we have developed to address this
problem is its seamless integration into the color glass condensate
framework, since it is formulated in terms of classical field
configurations (in fact, the resummation scheme we are using here was
obtained as a by-product of the formalism developed for proving the
initial state factorization of the large logs of $1/x_{1,2}$). As a
consequence, it can be adapted very easily to the Yang-Mills case, and
to the study of thermalization in high energy heavy ion collisions. A
crucial step in view of this application was performed in
\cite{DusliGV1}, where has been derived the spectrum of fluctuations
at early times (i.e. before the Glasma instabilities start to develop)
for gauge field fluctuations -- in a system of coordinates and a gauge
that are appropriate for numerical studies of the evolution of the
glasma fields.

Let us finally mention some possible connections with the idea
of ``eigenstate thermalization'' proposed by Srednicki in
\cite{Sredn1} (see also \cite{Deuts1,Jarzy1,RigolDO1}). This idea
follows Berry's conjecture~\cite{Berry1}, which states that the high
energy eigenstates of quantum systems whose classical counterpart is
chaotic have extremely complicated wave-functions that for all
practical purposes can be replaced by a random sum of plane waves
(with Gaussian distributed coefficients). Assuming that this
conjecture is satisfied, Srednicki proved that sufficiently inclusive
measurements on such a state would lead to results that agree with
thermal equilibrium. The main impediment to thermalization in a
quantum system would thus be the fact that the system is usually not
prepared in an energy eigenstate. For instance, for a system that is
initially in a coherent state, thermal features would only become
manifest once the various energy eigenstates superimposed in the
coherent state have become incoherent. In the problem we have
considered in the present paper and in \cite{DusliEGV1}, the initial
condition at $t=0$ appears to be closely related to the Wigner
distribution of a coherent state and the associated classical dynamics
is chaotic. Moreover, we have seen that the relaxation of the pressure
is due to the loss of coherence between the various initial
conditions.  It would thus be very interesting to study whether there
is a deeper connection between thermalization in quantum field theory
and the ideas advocated by Srednicki.

\section*{Acknowledgements}
We would like to thank J.P. Blaizot, K. Dusling, K. Fukushima,
Y. Hatta, K. Itakura, J. Liao, L. McLerran and R. Venugopalan for
useful discussions on the issues studied in this paper and on closely
related questions.

\appendix

\section{Evolution of the classical phase-space density}
\label{sec:classical}
In the figure \ref{fig:fk}, we have obtained a very good
fit of the occupation number at late times by a function of the form
\begin{equation}
f_\k=\frac{T}{\omega_\k-\mu}-\frac{1}{2}\; .
\label{eq:fk_class}
\end{equation}
This fit works except for the zero mode, that is over occupied with
respect to this distribution. We interpreted this distribution as the
classical approximation of a Bose-Einstein distribution, and the
$-1/2$ term was simply due to our definition of the occupation number.

However, it is also interesting to forget the underlying quantum
field theory we started from, and to consider in its own right the
classical problem by which it is approximated. This reformulation is
equivalent to solving the Liouville equation,
\begin{equation}
\partial_t{\cal F}_t+\{{\cal F}_t,{\cal H}\}=0\; ,
\end{equation}
given some Gaussian initial distribution. Therefore, if we adopt this
point of view, we just have a (large) collection of coupled classical
oscillators, and we follow their Hamiltonian flow in phase-space. In
this appendix, we discuss some aspects of this classical dynamical
system, that are relevant to the topics discussed in the rest of the
paper.

\subsection{Effective Hamiltonian}
Motivated by the observation of quasi-particles in the system, we
may assume that there is a transformation of the fields and their
conjugate momenta such that the Hamiltonian becomes a sum of
quasi-free harmonic oscillators coupled only by weak residual
interactions. In practice, this amounts to writing
\begin{eqnarray}
{\cal H}&=& \int d^3 \x\; \frac{1}{2}\Big(\dot\varphi^2+({\nabla}\varphi)^2\Big)
+\frac{g^2}{4!}\varphi^4\nonumber\\
&=&
\int d^3 \x\; \underbrace{\frac{1}{2}\Big(\dot\varphi^2+({\nabla}\varphi)^2+m^2\varphi^2\Big)}_{{\cal H}_0}
+\underbrace{\frac{g^2}{4!}\varphi^4-\frac{1}{2}m^2\varphi^2}_{{\cal H}^\prime_{\rm int}}\; .
\end{eqnarray}
So far, we have just added and subtracted a mass term by hand, and the
parameter $m^2$ is still arbitrary. In order to make the residual
interactions small, one can choose the mean field value for $m^2$,
\begin{equation}
m^2 = \frac{g^2}{2}\left<\varphi^2(x)\right>\; ,
\end{equation}
where the angle brackets denote an ensemble average\footnote{One can
  check numerically that this mean field expression of the mass is in
  very good agreement with the measured mass of the
  quasi-particles.}. The first part of this Hamiltonian can be
rewritten as a sum of independent harmonic oscillators by going to
Fourier space\footnote{Note that this and subsequent integrals over
  $d^3\k$ are strongly ultraviolet divergent. In all this appendix,
  one should have a lattice regularization in mind, so that the number
  of degrees of freedom in the system is finite.},
\begin{equation}
{\cal H}_0
=
\int\frac{d^3\k}{(2\pi)^3}\;
\underbrace{\frac{1}{2}\big|\dot\varphi_\k\big|^2
+
\frac{1}{2}\omega_k^2\big|\varphi_\k\big|^2}_{h_\k}\; ,
\end{equation}
where $\omega_\k\equiv (\k^2+m^2)^{1/2}$ and where $\varphi_\k$ is the
spatial Fourier transform of $\varphi$.

This decomposition of the classical Hamiltonian into elementary
harmonic oscillators plus residual interactions is a good starting
point to make connections with the study of the occupation number in
the previous sections. Indeed, it is easy to check that
eq.~(\ref{eq:fk}) is equivalent to
\begin{equation}
\frac{1}{2}+f_\k
=
\frac{\big<h_\k\big>}{V\omega_\k}\; .
\end{equation}
In other words, $\big<h_\k\big>$ is the occupancy of the mode $\k$
times $\omega_\k$ times the volume, plus a constant {\sl vacuum
  contribution} $V\omega_\k/2$. This means that one should find a
non-zero average value for $\big<h_\k\big>$ even in the vacuum --
actual particles in the mode $\k$ correspond to an excess of
$\big<h_\k\big>$ over $V\omega_\k/2$.

\subsection{Time evolution of the classical distribution}
\begin{figure}[p]
  \begin{center}
    \resizebox*{12.65cm}{!}{\rotatebox{-90}{\includegraphics{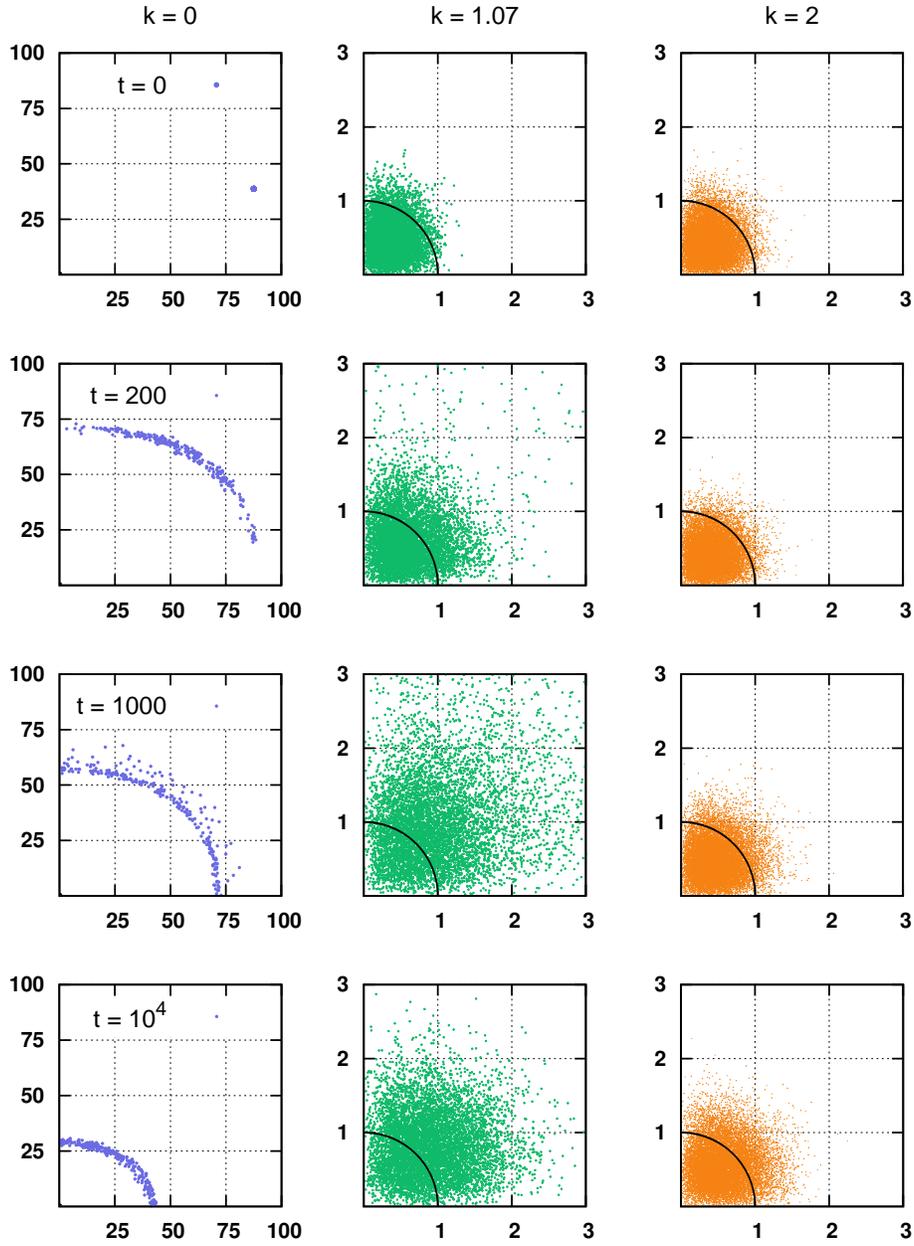}}}
  \end{center}
  \caption{\label{fig:6:phase-space}Phase-space density in the
    $(\sqrt{\omega_\k/V}|\varphi_\k|,|\dot\varphi_\k|/\sqrt{\omega_\k V})$
    plane. From left to right: $k=0$, $1.07$ (resonant mode), and
    $2$. From top to bottom: $t=0$, $200$, $10^3$ and $10^4$.
    Note the vastly different scale used for the zero mode (left
    column). The black circle represents the radius of the pure vacuum
    fluctuations.}
\end{figure}
In order to start the discussion, we first plot in the figure
\ref{fig:6:phase-space} the distribution ${\cal
  F}_t[\varphi,\dot\varphi]$ of the classical field configurations at
various stages of the evolution. Since we obviously cannot plot a
functional, we have represented several {\sl Fourier slices}: a slice
being defined as the distribution of configurations in the
plane\footnote{We rescale the Fourier components in this way so that
  the vacuum fluctuations look the same for all $\k$. A value of
  $\big<h_\k\big>$ proportional to $\omega_\k$ corresponds to a circle
  of fixed radius (of order unity) in this plane.}
$(\sqrt{\omega_\k/V}|\varphi_\k|,|\dot\varphi_\k|/\sqrt{\omega_\k
  V})$. In the figure \ref{fig:6:phase-space}, we have represented
this distribution for three values of $\k$: the zero mode, a mode in
the resonance band, and a mode at some higher momentum. At the initial
time, the zero mode of the fields is highly coherent, and its
distribution is concentrated around a single point. In contrast, all
the higher modes contain only fluctuations centered around the origin
$(0,0)$, with a width which is that of the vacuum fluctuations
(i.e. the width that gives $\big<h_\k\big>=\omega_\k/2$). At the next
time, $t=200$, the zero mode has decohered and now fills almost
uniformly a curve\footnote{This is not exactly a curve of constant
  $h_0$, as one can see from the fact that it is not a circle. This
  deviation from a circle is due to the fact that there is a large
  correction $g^2\varphi_0^4/4!$ coming from the interaction
  energy. Indeed, the zero mode, because of its large amplitude, is
  strongly interacting with itself.}  of constant energy. We also see
that the distribution of the resonant modes has expanded due to
parametric resonance, while the harder modes still have the same
distribution as the $t=0$ one. At later times, the expansion of the
resonant modes reaches a maximum and then subsides, while the
distribution of the hard modes begins to expand as
well. Simultaneously, the zero mode distribution shrinks slowly, while
remaining on a curve of constant energy (i.e. this seemingly constant
energy is in fact slowly decreasing, due to a transfer of energy from
the soft to the hard modes).

To summarize these observations, the Fourier modes that initially have
a large amplitude quickly decohere.  If there are many such modes
(instead of only one as in our numerical example), what happens then
depends on whether the classical dynamics is chaotic or on the
contrary integrable: in a chaotic system, all these modes mix and fill
uniformly an energy shell, while in the integrable case these modes
would evolve independently and cover an invariant torus of much
smaller dimension (for $N$ modes, an energy shell is a manifold of
dimension $2N-1$, while an invariant torus has dimension $N$ only).
On larger time scales, the energy carried by these initially large
modes decreases slowly, and is transferred to the higher modes whose
distribution expands as a consequence of this transfer.

\subsection{Asymptotic behavior}
What is then the asymptotic distribution ${\cal
  F}_\infty[\varphi,\dot\varphi]$ of these classical fields? Let us
assume first that the only quantity that is invariant under the
Hamiltonian flow is the energy itself\footnote{The total spatial
  momentum of the system is also conserved, but it does not play any
  role here because we analyze the problem in the rest frame of the
  system.}. From the Liouville equation, it is clear that the
asymptotic distribution must have a vanishing Poisson bracket with the
Hamiltonian. This property is satisfied by any distribution that
depends on $\varphi,\dot\varphi$ only through the Hamiltonian ${\cal
  H}[\varphi,\dot\varphi]$.  Such an asymptotic form for the
distribution ${\cal F}_\infty$ has strong implications on the average
value of $\left<h_\k\right>$. Indeed, if we can neglect the
interaction energy (e.g. assuming that most of the interaction energy
can be absorbed into an effective mass), then a distribution that
depends only on ${\cal H}$ implies the equipartition of the energy
among the modes,
\begin{equation}
\left<h_\k\right>=\mbox{const (independent of $\k$)}\; .
\label{eq:equipartition}
\end{equation}
Note that this conclusion holds no matter what is the precise form of
the function of ${\cal H}$ that ${\cal F}_\infty$ is equal to. When
equipartition is in the classical phase-space is reached, the
occupation number becomes of the form:
\begin{equation}
f_\k = \frac{\mbox{const}}{\omega_\k}-\frac{1}{2}\; .
\label{eq:fk-equi}
\end{equation}

\subsection{Asymptotic behavior with constraints}
\begin{figure}[htbp]
  \begin{center}
    \resizebox*{8cm}{!}{\rotatebox{-90}{\includegraphics{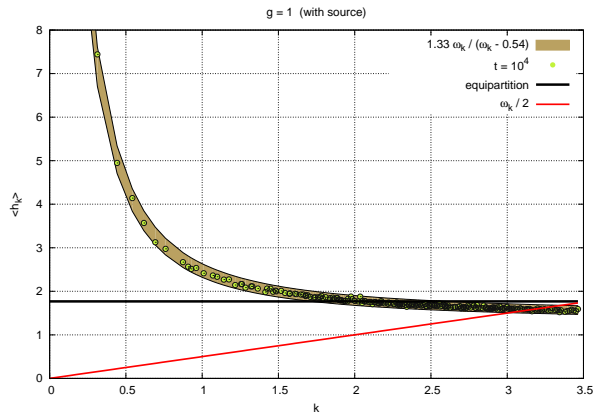}}}
  \end{center}
  \caption{\label{fig:edist_j} Average value $\big<h_\k\big>$ of the
    energy in the mode $\k$. The brown band is a fit of the form
    $T\omega_\k/(\omega_\k-\mu)$, and its width reflects the expected
    statistical error in our Monte-Carlo simulation. The horizontal
    line shows the expected result if equipartition is reached. The
    red line represents the zero point energy $\omega_\k/2$.}
\end{figure}
When we try to fit the late time curves of the figure \ref{fig:fk} by
an expression of the form (\ref{eq:fk-equi}), that lacks the chemical
potential $\mu$, the results are not good. This is best seen by
plotting $\left<h_\k\right>$ (see the figure \ref{fig:edist_j}), since
eq.~(\ref{eq:fk-equi}) would correspond to an horizontal line in this
plot. In contrast, the much better fit of the figure \ref{fig:fk}
could be explained if the average value of $h_\k$ at late times was of
the form
\begin{equation}
\left<h_\k\right>= \mbox{const}\; \frac{\omega_\k}{\omega_\k-\mu}\; .
\label{eq:equi-N}
\end{equation}
This is obvious in the figure \ref{fig:edist_j}, where we
obtain a good fit to $\big<h_\k\big>$ by an expression of this form.

It is possible to understand eq.~(\ref{eq:equi-N}) if one assumes that
the Hamiltonian flow not only conserves energy (exactly), but also (to
a good degree of approximation) the following quantity\footnote{The
  average $\left<{\cal N}\right>$ in the classical dynamical system is
  the counterpart of the number of quasi-particles in the quantum
  theory.}
\begin{equation} {\cal N}[\varphi,\dot\varphi]\equiv \int
  \frac{d^3\k}{(2\pi)^3}\; \frac{h_\k}{\omega_\k}\; .
\label{eq:N-class}
\end{equation}
It is obvious that this quantity has a vanishing Poisson bracket with
the free quasi-particle part of the Hamiltonian ${\cal H}_0$, and that
only the residual quasi-particle interactions ${\cal H}^\prime_{\rm
  int}$ can possibly make this quantity change. Thus, if the residual
interactions are weak, this quantity should indeed be approximately
conserved.  If both ${\cal H}$ and ${\cal N}$ are invariants, then the
most general asymptotic solution of the Liouville equation must depend
on $\varphi,\dot\varphi$ only through some combination of the form
${\cal H}-\mu{\cal N}$. For any distribution of this form,
equipartition is replaced by
\begin{equation}
\left<h_\k-\mu\frac{h_\k}{\omega_\k}\right>=\mbox{const (independent of $\k$)}\; ,
\label{eq:hk-mu}
\end{equation}
which leads to eq.~(\ref{eq:equi-N}). Therefore, in the classical
dynamical system, the quasi conservation of ${\cal N}$ is what
explains that we observed an energy distribution that differs
considerably from the naive equipartition.

Note that when $k$ is large, eq.~(\ref{eq:hk-mu}) is equivalent to
$\left<h_\k\right>=\mbox{const}$. This implies that the constant in
the right hand side must be positive. Considering now $\k=0$, this
imposes $\mu\le m$. Inserting now eq.~(\ref{eq:hk-mu}) in the
definition of ${\cal N}$,
\begin{equation} \left<{\cal
    N}\right>=\int\frac{d^3\k}{(2\pi)^3}\;\frac{\mbox{const}}{\omega_\k-\mu}\;,
\end{equation}
we see that it increases with $\mu$, and reaches a finite (because the
singularity at $\k=0$ is integrable) maximum when $\mu=m$. However, it
may happen that the initial value of $\big<{\cal N}\big>$ is larger
than this maximum. In this situation, the $\big<h_\k\big>$'s must be
altered in order to accommodate this excess, but in such a way that
eq.~(\ref{eq:hk-mu}) remains valid. Any modification of
$\big<h_\k\big>$ for $\k\not=0$ will violate eq.~(\ref{eq:hk-mu}), so
the only possibility is to modify $\big<h_0\big>$. Then, we see that
the only way to change $\big<h_0\big>$ without violating
eq.~(\ref{eq:hk-mu}) is to have $\mu=m$. Thus, when there is an
excess of $\big<{\cal N}\big>$, the equilibrium value of
$\left<h_\k\right>$ takes the form
\begin{equation}
\left<h_\k\right>= A\,\delta(\k)+B\,\frac{\omega_\k}{\omega_\k-m}\; .
\end{equation}
This explains why we found a chemical potential whose value is very
close to the mass of the quasi-particles (within statistical errors).
This phenomenon can be seen as an analogue in classical Hamiltonian
dynamics of Bose condensation in quantum mechanics.

\subsection{Do vacuum fluctuations thermalize?}
The fact that the Liouville evolution makes the distribution ${\cal
  F}_t$ evolve towards a classical thermal equilibrium leads to an
interesting question: does the same happen if in our original problem
there is no source coupled to the quantum fields, i.e. for pure vacuum
fluctuations?  In this case, the initial distribution is
\begin{equation}
{\cal F}_0[\varphi,\dot\varphi]
=
\exp\Bigg[-\int\frac{d^2\k}{(2\pi)^3}\frac{|\dot\varphi_\k|^2+k^2|\varphi_\k|^2}{k}
\Bigg]\; ,
\label{eq:vac-fluct}
\end{equation}
corresponding to $\left<h_\k\right>=k/2$. The consistency of our
approach requires that the vacuum does not change over time. However,
since eq.~(\ref{eq:vac-fluct}) is not a function of ${\cal H}$, our
previous considerations suggest that equipartition may also occur if
we start from the pure vacuum fluctuations given by
eq.~(\ref{eq:vac-fluct}).
\begin{figure}[htbp]
  \begin{center}
    \resizebox*{8cm}{!}{\rotatebox{-90}{\includegraphics{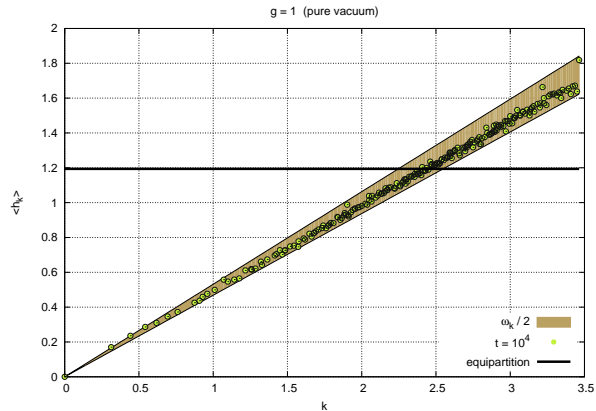}}}
  \end{center}
  \caption{\label{fig:edist_vac}Average value of the energy $h_\k$ in
    the mode $\k$ for pure vacuum fluctuations. The center of the
    brown band is $\omega_\k/2$, and its width reflects the expected
    statistical error in our Monte-Carlo simulation). The horizontal
    line corresponds to equipartition.}
\end{figure}
We have computed $\left<h_\k\right>$ numerically for the pure vacuum
case, by starting from the distribution of eq.~(\ref{eq:vac-fluct})
and by evolving the corresponding classical field configurations also
to a large time $t=10^4$. As shown in the figure \ref{fig:edist_vac},
this computation shows no sign of equipartition of the vacuum energy,
indicating that the vacuum is stable in our framework.

This is at first sight an intriguing result. Indeed, how does the
purely classical Liouville equation know that the distribution of
eq.~(\ref{eq:vac-fluct}) represents the ground state of the
corresponding quantum theory? In a sense, the Liouville equation seems
to know more than what its ``classical'' qualifier might
suggest. This result can be qualitatively explained by recalling a
formal connection that exists between quantum mechanics and the
classical Liouville equation. In quantum mechanics, a system can be
described by a density operator $\hat\rho$, whose evolution is driven
by the Von Neumann equation,
\begin{equation}
i\partial_t\hat\rho+[\hat\rho,H]=0\; .
\label{eq:vonneumann}
\end{equation}
It was noted by Wigner and Moyal that this quantum mechanical equation
can be formulated equivalently as an evolution in the classical
phase-space (see \cite{Polko1} for a recent review). To do this, one
should first introduce the Wigner distribution associated to the
density operator $\hat\rho$. For a single degree of freedom, this is
defined as
\begin{equation}
W(q,p)\equiv \int dx\; e^{ipx} \; \big<q+x/2\big|\hat\rho\big|q-x/2\big>\; . 
\end{equation}
The Von Neumann equation can then be shown to be equivalent to the
Moyal equation for $W$,
\begin{equation}
\partial_t W+\{\{W,{\cal H}\}\}=0\; ,
\label{eq:moyal}
\end{equation}
where $\{\{\cdot,\cdot\}\}$ is the Moyal bracket, obtained as the
Wigner transform of the commutator. At this stage, one has a
formulation of quantum mechanics that involves only quantities defined
on the classical phase-space (the quantum mechanical aspects are
hidden in the fact that the Moyal bracket depends on $\hbar$).  The
connection to the classical Liouville equation arises via the
following two properties,
\begin{itemize}
\item[{\bf i.}] the Moyal bracket $\{\{\cdot,{\cal H}\}\}$ is equal to
  the Poisson bracket $\{\cdot,{\cal H}\}$ if the Hamiltonian ${\cal
    H}$ is quadratic in the coordinates and momenta,
\item[{\bf ii.}] for any Hamiltonian, one has
  \begin{equation}
    \{\{A,B\}\} = \{A,B\}+{\cal O}(\hbar^2)\; .
  \end{equation}
\end{itemize}
If the quantum system is in its ground state $\big|0\big>$, its
density operator $\hat\rho_0\equiv \big|0\big>\big<0\big|$ is
invariant under the Von Neumann equation. Equivalently, the
corresponding Wigner distribution $W_0$ is invariant under the Moyal
equation. Since ${\cal F}_0$ is the Wigner distribution of the same
vacuum state, it is thus natural that it is left invariant by the
Liouville evolution, since it is the $\hbar\to 0$ limit of the
invariant Moyal evolution.

\section{Liouville equation}
\label{app:liouville} In this appendix, we recall the derivation of
Liouville's equation and some of its well known properties.

\subsection{Hamilton's equations}
In this appendix, we denote generically the canonical coordinates by
${\bs Q}$, the corresponding canonical momenta by $\P$, and the
Hamiltonian by ${\cal H}$. Hamilton's equations read
\begin{equation}
\dot{\Q} = \frac{\partial{\cal H}}{\partial \P}
\;,\qquad
\dot{\P} = -\frac{\partial{\cal H}}{\partial \Q}\; .
\end{equation}
Let us introduce a few useful notations~:
\begin{equation}
\X \equiv \begin{pmatrix}\Q\\\P\end{pmatrix}\;,\qquad
{\bs\nabla}\equiv
\begin{pmatrix}\partial/\partial\Q\\\partial/\partial\P\end{pmatrix}\;,\qquad
{\bs V}\equiv \begin{pmatrix}
{\partial{\cal H}}/{\partial \P}\\
-{\partial{\cal H}}/{\partial \Q}
 \end{pmatrix}\; .
\end{equation}
Hamilton's equations can now be rewritten in the following compact form
\begin{equation}
\dot\X = {\bs V}\; .
\end{equation}
This notation is suggestive of the fact that ${\bs V}$ is the velocity
field induced on phase-space by the Hamiltonian. A crucial property of
Hamiltonian flows is that they are incompressible,
\begin{equation}
{\bs\nabla}\cdot{\bs V}=0\; .
\label{eq:appF:incompress}
\end{equation}
This identity is the essence of Liouville's equation.

\subsection{Liouville's equation}
Consider an ensemble of such dynamical systems, all described by the
same Hamiltonian ${\cal H}$. At the time $t$, their distribution in
phase-space is described by a density ${\cal F}_t[\Q,\P]$, and we wish
to derive an equation that describes the time evolution of this
distribution. Naturally, the number of systems in the ensemble is not
changing since each system evolves independently of the others. Thus,
${\cal F}_t$ is the density for a locally conserved quantity. We have
seen before that each point in phase-space moves with the velocity
${\bs V}$. Therefore, we can write a continuity equation that expresses
this conservation,
\begin{equation}
\partial_t{\cal F}_t+{\bs\nabla}\cdot\left({\cal F}_t{\bs V}\right)=0\; .
\label{eq:appF:liou1}
\end{equation}
By using eq.~({\ref{eq:appF:incompress}}), we can rewrite this
equation as\footnote{In the form $D_t{\cal F}_t=0$, the Liouville
  equation is equivalent to the {\sl Liouville theorem}, that states
  that ${\cal F}_t$ is constant along the Hamiltonian flow lines.}
\begin{equation}
\Big[\underbrace{\partial_t+{\bs V}\cdot{\bs\nabla}}_{D_t}\Big]{\cal F}_t=0\; .
\label{eq:appF:liou2}
\end{equation}
Note that the term ${\bs V}\cdot{\bs\nabla}{\cal F}_t$ is nothing but
the Poisson bracket $\{{\cal F}_t,{\cal H}\}$. Thus, we see that both
eqs~(\ref{eq:appF:liou1}) and (\ref{eq:appF:liou2}) are equivalent to
the usual form of the Liouville equation,
\begin{equation}
\partial_t{\cal F}_t +\{{\cal F}_t,{\cal H}\}=0\; .
\label{eq:appF:liou3}
\end{equation}
These alternate forms of the Liouville equation are very useful,
because they recall us its origin as the continuity equation for the
density of systems in phase-space, and because they make some
properties of Hamiltonian flows more transparent -- in particular
thanks to the {\sl flow derivative} $D_t$ introduced in
eq.~(\ref{eq:appF:liou2}).

\subsection{Basic properties}
In this section, we derive some elementary properties of Liouville's
equation. To state some of these properties, it will be useful to
denote $[d{\bs\Gamma}]$ the measure on phase-space.

The most elementary property,
\begin{equation}
\int[d{\bs\Gamma}]\;{\cal F}_t = \mbox{const}\; , 
\label{eq:appF:cons1}
\end{equation}
is in fact just the integral version of Liouville's equation
itself. It is simply another way of stating that the number of systems
in the ensemble does not change over time.  Note that since ${\cal
  F}_t$ is constant along the flow lines, the same is true of any
local function $F({\cal F}_t)$. Thus, eq.~(\ref{eq:appF:cons1}) can be
generalized by replacing under the integral ${\cal F}_t$ by any
function $F({\cal F}_t)$. A similar statement can be made about the
energy if we note that\footnote{This is due to $\partial_t{\cal H}=0$
  and ${\bs V}\cdot{\bs\nabla}{\cal H}=\{{\cal H},{\cal H}\}=0$.}
\begin{equation}
D_t {\cal H} =0\; .
\end{equation}
(which means that the Hamiltonian does not vary if we follow the
flow). Then, by multiplying eq.~(\ref{eq:appF:liou2}) by ${\cal H}$,
and by integrating over phase-space, we get
\begin{equation}
\int[d{\bs\Gamma}]\;{\cal F}_t\,{\cal H} = \mbox{const}\; .
\label{eq:appF:cons2}
\end{equation}
This equation simply says that the total energy of our ensemble of
systems is conserved.

\bibliographystyle{unsrt}

\end{document}